\documentclass[12pt]{article}

\usepackage{graphicx}
\usepackage{amsmath,amssymb} 
\usepackage{latexsym}
\usepackage{mathrsfs}
\usepackage{bbold}
\usepackage{color}
\usepackage{slashed}
\definecolor{red}{rgb}{1,0,0}
\usepackage{cancel}
\usepackage{comment}
\usepackage{cite,mathtools}
\usepackage{caption}
\usepackage{subcaption}
\definecolor{red}{rgb}{1,0,0}

\newcommand{\be}{\begin{equation}}
\newcommand{\ee}{\end{equation}}
\newcommand{\bee}{\begin{equation*}}
\newcommand{\eee}{\end{equation*}}
\newcommand{\bea}{\begin{eqnarray}}
\newcommand{\eea}{\end{eqnarray}}
\newcommand{\bean}{\begin{eqnarray*}}
\newcommand{\eean}{\end{eqnarray*}}

\def\bea{\begin{eqnarray}} \def\eea{\end{eqnarray}}
\def\be{\begin{equation}} \def\ee{\end{equation}}

\newcommand{\sgn}{\text{sign}}

\newcommand{\promille}{%
  \relax\ifmmode\promillezeichen
        \else\leavevmode\(\mathsurround=0pt\promillezeichen\)\fi}
\newcommand{\promillezeichen}{%
  \kern-.05em%
  \raise.5ex\hbox{\the\scriptfont0 0}%
  \kern-.15em/\kern-.15em%
  \lower.25ex\hbox{\the\scriptfont0 00}}

\newcommand{\ir}{{_{\rm IR}}}
\newcommand{\uv}{{_{\rm UV}}}

\usepackage[colorlinks,linkcolor=black,citecolor=blue,urlcolor=blue,linktocpage]{hyperref}

\makeatletter
\def\section{\@startsection {section}{1}{\z@}{-3.5ex plus -1ex minus
 -.2ex}{2.3ex plus .2ex}{\large\bf}}
\def\subsection{\@startsection{subsection}{2}{\z@}{-3.25ex plus -1ex
minus -.2ex}{1.5ex plus .2ex}{\normalsize\bf}}
\makeatother
\makeatletter

\@addtoreset{equation}{section}

\makeatother

\begin{document}

\setcounter{page}{0}
\thispagestyle{empty}

\begin{flushright}
DESY 17-056
\end{flushright}

\vskip 8pt

\begin{center}
{\bf \LARGE {QCD-induced Electroweak Phase Transition}}
\end{center}

\vskip 12pt

\begin{center}
 { \bf  Benedict von Harling$^a$ and G\'eraldine Servant$^{a,b}$}
 \end{center}

\vskip 14pt

\begin{center}
\centerline{$^{a}${\it DESY, Notkestrasse 85, 22607 Hamburg, Germany}}
\centerline{$^{b}${\it II.~Institute of Theoretical Physics, University of Hamburg, 22761 Hamburg, Germany}}

\vskip .3cm
\centerline{\tt benedict.von.harling@desy.de, geraldine.servant@desy.de}
\end{center}

\vskip 10pt

\begin{abstract}
\vskip 3pt
\noindent

Phase transitions associated with nearly conformal dynamics are known to lead to significant supercooling.
A notorious example is the phase transition in Randall-Sundrum models or their CFT duals. 
In fact, it was found that the phase transition in this case is first-order and the tunneling probability for the radion/dilaton is so small that the system typically remains trapped in the false vacuum and the phase transition never completes. The universe then keeps expanding and cooling. Eventually the temperature drops below the QCD scale. We show that the QCD condensates which subsequently form give an additional contribution to the radion/dilaton potential, an effect which had been ignored so far. This significantly reduces the barrier in the potential and allows the phase transition to complete in a substantially larger region of parameter space. Due to the supercooling, electroweak symmetry is then broken simultaneously. 
This class of models therefore naturally leads to an electroweak phase transition taking place at or below QCD temperatures, with interesting cosmological implications and signatures.

\end{abstract}

\newpage

\tableofcontents

\vskip 13pt

\newpage

\section{Introduction}

The nature of the electroweak phase transition is still weakly constrained experimentally and many possibilities remain open, in particular when the scalar sector of the theory is extended.
There has been growing interest lately in the possibility of a strong first-order electroweak phase transition, not only because of its  relevance for baryogenesis but also because it is a potential source of gravitational waves detectable at LISA \cite{Caprini:2015zlo}. The majority of the literature has focused on polynomial potentials for the Higgs and associated scalar fields.

On the other hand, if the Higgs is part of an approximately conformal sector, electroweak symmetry breaking is tied to the breaking of conformal invariance. The electroweak phase transition is then governed by a nearly conformal potential. This scenario has very interesting cosmological properties as such potentials generically lead to large amounts of supercooling, see e.g.~refs.~\cite{Witten:1980ez,Buchmuller:1990ds,Espinosa:2008kw,Konstandin:2011dr,Jaeckel:2016jlh,Marzola:2017jzl,Iso:2017uuu}. This may in particular delay the phase transition to temperatures near the QCD scale or below. When this happens, QCD confines and gluons and quarks form condensates. 
The main motivation of this paper is to study if these condensates can subsequently trigger the breaking of conformal invariance and thereby induce the electroweak phase transition.

For illustration, we work with the 5D Randall-Sundrum (RS) model \cite{Randall:1999ee}. Via the gauge-gravity duality, this and related constructions \cite{Agashe:2004rs,Contino:2003ve} are dual to composite Higgs models with partial compositeness, where the Higgs arises from a nearly conformal sector \cite{Contino:2010rs,Contino:2006nn,Panico:2015jxa}.
The RS model has been a very popular solution to the hierarchy problem. In addition, it provides a framework to address the flavour puzzle in the standard model \cite{Gherghetta:2000qt,Huber:2000ie,Gherghetta:2010cj}. The distance between the UV and IR brane in the RS model corresponds to a scalar field, the so-called radion.  This field maps to the dilaton of the dual CFT, the pseudo-Nambu-Goldstone boson of broken scale invariance.  
It can obtain a potential for example by means of the Goldberger-Wise mechanism \cite{Goldberger:1999uk} which stabilizes the inter-brane distance, or equivalently, triggers the breaking of conformal invariance. The corresponding potential is nearly conformal and has the form
\be
V(\mu) \, = \, \mu^4 \times f(\mu^{\epsilon}) \, ,
\ee
where $f(\mu^{\epsilon})$ with $|\epsilon| \ll 1$ is a very slowly-varying function of $\mu$. Due to this form, the potential is very shallow with extrema which are far separated from each other in field space. A potential of the Goldberger-Wise type in particular typically has a barrier at very small field values which separates the origin from the minimum. The dilaton/radion then needs to tunnel through this barrier during the phase transition which leads to the breaking of conformal invariance, or the stabilisation of the inter-brane distance, in the early universe (for earlier studies of this phase transition see \cite{Creminelli:2001th,Randall:2006py,Nardini:2007me,Konstandin:2010cd,Konstandin:2011dr,Hassanain:2007js,Dillon:2017ctw,Bunk:2017fic}). However, the vast distance between the extrema suppresses the tunneling rate. The phase transition can therefore typically not complete and the field instead remains stuck in the wrong vacuum. This is the origin of the supercooling that we have mentioned earlier.

The shallowness of the potential, however, also means that corrections to the potential at small field values can have a big impact. As we will show in this paper, such corrections can in particular arise from the QCD condensates which form when the temperature drops to the QCD scale. In the RS model, the gluon is a 5D bulk field. The gauge coupling of the 4D massless mode which we identify with the gluon is then affected by the length of the extra dimension and thus by the radion.  In the dual CFT, on the other hand, the gluon gauges an $SU(3)$ symmetry of the CFT. The resulting contribution to the QCD $\beta$-function decouples when the CFT confines. This leads to a dependence of the QCD coupling on the confinement scale of the CFT and thus on the dilaton. The scale where QCD becomes strongly coupled and itself confines then also depends on the radion/dilaton. Via the gluon and quark condensates, this gives an additional contribution to the potential which can change its shape near the barrier. This can in turn significantly enhance the tunneling probablility, allowing the phase transition to complete. 

The Higgs is localized towards the IR brane of the RS model which in the dual CFT corresponds to the Higgs being a composite state. The fact that in this scenario the temperature during the phase transition of the radion/dilaton is below the QCD scale, means that temperature corrections to the Higgs potential are negligible. Electroweak symmetry will therefore generically be broken simultaneously and QCD thus induces the electroweak phase transition. This potential dramatic impact on the nature of the electroweak phase transition was overlooked in the previous literature on RS and composite Higgs models.
A fascinating application would be to use the strong $CP$-violation from the QCD axion, which would be large during such a phase transition near QCD temperatures, as the source of $CP$-violation for cold baryogenesis \cite{Servant:2014bla}. The output of this work will give strong support for this possibility.

The outline of the paper is as follows. We begin with a review of the Goldberger-Wise potential for the radion in sec.~\ref{GWReview}. In sec.~\ref{sec:QCDIR}, we then show how the QCD confinement scale depends on the vacuum expectation value of the radion. Sec.~\ref{sec:QCDcont} presents the new contribution to the radion potential from the QCD condensate. In sec.~\ref{sec:PT}, we then review the RS phase transition, first without the QCD effect and then showing the impact of the QCD condensate on the nature of the phase transition. 
We comment on cosmological implications and experimental signatures in sec.~\ref{sec:cosmoimpli} and conclude in sec.~\ref{sec:conc}.

\section{The radion potential in Randall-Sundrum models}
\label{GWReview}

We begin with a review of the important properties of the Randall-Sundrum model from which the nearly-conformality of the scalar potential originates.
The geometry is that of a slice of AdS$_5$ space with metric
\be 
ds^2 \, = \, e^{-2 k y} \, \eta_{\mu \nu} dx^\mu dx^\nu \, - \, dy^2 \, ,
\ee
where $k\sim {\cal O}(M_{\rm Pl})$ is the AdS$_5$ curvature.
The slice is bounded by two branes at $y=0$ and $y=y_\ir$ to which we refer as the UV and IR branes, respectively. It can be obtained either from an 
orbifold or directly from an interval. In either case, we shall restrict the coordinate to the interval $[0,y_\ir]$ here and below. The size $y_\ir$ of the extra dimension can be stabilized by means of the Goldberger-Wise mechanism \cite{Goldberger:1999uk}. To this end,  a bulk scalar is introduced,
\be
\label{actionGW}
 S \, \supset \, \int d^5 x \,\sqrt{g} \, \Bigl(\frac{1}{2} \partial_A \phi \, \partial^A \phi  -  \frac{m_\phi^2}{2} \phi^2  -  \delta(y) \, V_\uv  -  \delta(y-y_\ir) \, V_\ir \Bigr) 
\ee
with boundary potentials
\be
V_\uv \, = \,  \lambda_\uv (\phi^2 - v^2_\uv k^3)^2\, , \quad \quad \quad V_\ir \, = \, \lambda_\ir (\phi^2 - v^2_\ir k^3)^2 \, .
\label{BoundaryPotentials}
\ee
These trigger a vacuum expectation value (VEV) for the scalar with profile along the extra dimension given by
\be
\label{vevprofile}
\langle \phi \rangle \, = \, A \, k^{3/2}  e^{(4+\epsilon) k y} \, + \, B \,k^{3/2}  e^{-\epsilon k y} \, ,
\ee
where 
\be
\epsilon \equiv \sqrt{4+m_\phi^2/k^2}-2 \, .
\ee
The mass of a scalar in AdS$_5$ can be tachyonic, $m_\phi^2 \geq -4 k^2$ according to the Breitenlohner-Freedman bound \cite{Breitenlohner:1982bm}, and $\epsilon$ can thus be both positive and negative. 
The integration constants $A$ and $B$ are determined by the boundary conditions which depend on the boundary potentials. In the limit of large couplings $\lambda_\ir$ and $\lambda_\uv$, one finds
\begin{align} 
\label{leadingA}
A \, & = \, \frac{v_\ir - v_\uv e^{-\epsilon ky_\ir}}{e^{(4+\epsilon)k y_\ir}-e^{-\epsilon ky_\ir}} \, \simeq \, v_\ir  \, e^{-(4+\epsilon)k y_\ir} \, - \, v_\uv  \,e^{-(4+2\epsilon)k y_\ir}\, ,\\
B \, & = \, v_\uv -A \, \simeq \, v_\uv \, .
\label{leadingB}
\end{align} 
The scalar VEV and its contribution to the energy-density thus depends on the size of the extra dimension.
Integrating over the extra dimension then leads to the effective 4D potential
\be 
\label{RadionPotential}
V_{\rm GW}(\mu) \, \simeq \,  \mu^4 \left[ (4+ 2 \epsilon)\, \left(v_\ir \, - \, v_\uv \left( \frac{\mu}{k} \right)^\epsilon  \right)^2\, - \, \epsilon  \, v_\ir^2 \, + \, \delta \right] \, ,
\ee
where
\be 
\mu \, \equiv \, e^{-k y_\ir} k \,. 
\ee
We will refer to this field as the radion. In the potential, we have neglected a constant piece but included an additional contribution $\delta \, \mu^4$. The latter arises if the IR brane tension $T$ is detuned from the value which is required to obtain a static solution in the Randall-Sundrum model without radion stabilization, $T=-24 M_5^3 k +  \delta \, k^4 $ with $M_5$ being the 5D Planck scale. Due to various loop corrections on the IR brane, $\delta$ is generically expected to be nonzero. 

Provided that $-(4+\epsilon) v_\ir^2 < \delta < (\epsilon+\epsilon^2/4) v_\ir^2$, the above potential has a global minimum and one maximum at
\be 
\label{GWminmax}
\mu_{\rm min,max} \, \simeq \, k \, \left(\frac{v_\ir}{v_\uv} \right)^{1/\epsilon} \, X_{\rm min,max}^{1/\epsilon} \,\, ,
\ee
where
\be 
X_{\rm min,max} \, \equiv \,\left(1 \, + \, \frac{\epsilon}{2} \right)^{-1} \left(1 \, + \, \frac{\epsilon}{4} \, \pm \, \frac{\sgn(\epsilon)}{2} \sqrt{\epsilon+\frac{\epsilon^2}{4} - \frac{\delta}{v_\ir^2}} \right)\, .
\ee
The radion can then be stabilized at hierarchically small values $\mu \ll k$ with an order-one ratio $v_\ir/v_\uv$ if $|\epsilon| \ll 1$.  
Note that $\mu_{\rm max} < \mu_{\rm min}$ and the maximum is thus a barrier which separates the origin from the minimum. As we discuss in more detail in sec.~\ref{sec:PT}, in the early universe the radion needs to transition from the origin to the minimum of the potential. The barrier therefore means that this phase transition is first-order. On the other hand, for $\delta < -(4+\epsilon) v_\ir^2$ the barrier disappears if $\epsilon >0$, while the potential then only has a minimum at the origin if $\epsilon <0$. Both the barrier and the minimum disappear for $\delta > (\epsilon+\epsilon^2/4) v_\ir^2$. One thus necessarily needs a nonvanishing $\delta$ if $\epsilon$ is negative. 
We can use the above relation to trade $v_\uv$ for $\mu_{\rm min}$. The potential then reads
\be 
\label{RadionPotential2}
V_{\rm GW}(\mu) \, = \,  \mu^4 \, v_\ir^2  \left[ (4+ 2 \epsilon)\, \left(1 \, - \,  X_{\rm min} \left( \frac{\mu}{\mu_{\rm min}} \right)^\epsilon \right)^2\, - \, \epsilon   \, + \, \frac{\delta}{v_\ir^2} \right] \, .
\ee

In the derivation so far we have tacitly assumed that the scalar VEV does not deform the geometry. As we discuss in the appendix, for $\epsilon>0$ this is fulfilled provided that
\be
\label{BackreactionConstraint}
\frac{v_\ir}{N} \, \ll \, \min\left[ \frac{1}{2 \pi X_{\rm min}} \sqrt{\frac{3}{\epsilon}}\left(\frac{\mu_{\rm min}}{k} \right)^\epsilon \, , \, \frac{\sqrt{3}}{4 \pi}\right] \, ,
\ee
where 
\be 
N \, \equiv \, 4 \pi \left(\frac{M_5}{k}\right)^{3/2} \, .
\ee
Using this parameter is motivated by the AdS/CFT correspondence which suggests that the gauge theory which is dual to the Randall-Sundrum model has $\mathcal{O}(N)$ colors (the precise prefactor is undetermined; the prefactor in the definition above arises for the gauge theory dual to type IIB string theory on AdS$_5\times S^5$). For later use, we note that the ratio $M_5/k$ and thus $N$ is restricted by the requirement that we can neglect higher powers of the Ricci scalar compared to the Einstein-Hilbert term in the action. Estimating the coefficients of these terms from naive dimensional analysis, this gives the condition \cite{Agashe:2007zd}
\be 
\label{Nconstraint}
N \, \gtrsim \,  \frac{4 \cdot \,5^{3/4}}{\sqrt{3 \pi}} \, .
\ee

Let us next consider the case $\epsilon<0$. As we discuss in more detail in the appendix, in this case there is always a region around the origin in the radion potential for which the backreaction of the Goldberger-Wise scalar on the geometry can not be neglected. It may still be possible to reliably analyse the phase transition if this region is sufficiently small. Nevertheless, we will focus on the case $\epsilon>0$ in this paper, which is enough for our purpose. 

Together with its kinetic term \cite{Csaki:1999mp,Goldberger:1999un}, the 4D action for the radion reads
\be 
\label{eq:kinetictermradion}
S \, \supset \, \int d^4x \, \left(\frac{3 N^2}{4 \pi^2} \left(\partial_\rho \mu \right)^2 \, + \, V_{\rm GW}(\mu) \right)\, .
\ee
The field $\mu$ is thus not canonically normalized. We will nevertheless continue to work with $\mu$ since it sets the mass scale of the Kaluza--Klein (KK) modes. The minimum $\mu_{\rm min}$ of its potential is therefore directly constrained by collider and flavour experiments and electroweak precision tests. We will use $\mu_{\rm min}=2.5\,$TeV throughout this paper.\footnote{With a custodial symmetry in the bulk, the bound from electroweak precision tests is $\mu_{\rm min}> 1.9 \,$TeV \cite{Malm:2013jia,Bauer:2016lbe}. An additional strong constraint arises from $CP$-violation in $K-\bar{K}$-mixing. This can be satisfied by either a larger $\mu_{\rm min}$ or an accidental cancellation of order $5 - 10\%$ \cite{Malm:2013jia,Bauer:2016lbe}. Alternatively, the relevant process can be suppressed by extending the QCD gauge group in the bulk \cite{Bauer:2011ah} or by coupling the Goldberger-Wise scalar to the bulk fermions \cite{vonHarling:2016vhf}. }

The potential (\ref{RadionPotential2}) is the basis for most studies of the phase transition in RS models \cite{Creminelli:2001th,Randall:2006py,Nardini:2007me,Konstandin:2010cd,Konstandin:2011dr,Hassanain:2007js,Dillon:2017ctw,Bunk:2017fic}. 
The position of the barrier and the shape of the potential in the vicinity of the barrier control the size of the tunneling action. On the other hand, for small $\epsilon$ as needed to explain the hierarchy between the electroweak (EW) and Planck scales, the position of the barrier is close to the origin of the potential. Any corrections of the potential 
at scales much smaller than the IR scale (i.e.~of order TeV) can therefore have a strong impact on the phase transition dynamics. The  purpose of this work is to consider the corrections from QCD confinement which were ignored so far. We will see that they substantially improve the tunneling probability and open up parameter space for a viable cosmology.

\section{Dependence of the QCD scale on the radion in Randall-Sundrum models}
\label{sec:QCDIR}

We next review relevant aspects of QCD in an RS model and then discuss how this leads to a dependence of the QCD scale on the radion. 
The action for QCD in an RS model reads
\be
\label{QCDaction}
S \, \supset \, \int d^5 x \, \sqrt{g} \, \left(-\frac{1}{4 g_5^2} G_{M N} G^{M N} \, - \, \delta(y) \, \frac{\tau_\uv }{4} G_{\mu \nu} G^{\mu \nu}\, - \, \delta(y-y_\ir) \, \frac{\tau_\ir }{4} G_{\mu \nu} G^{\mu \nu} \right) \, ,
\ee
where $G_{MN}$ and $g_5$ are the field strength and coupling of QCD in 5D and we have allowed for localized kinetic terms on the two branes. Performing a KK decomposition and integrating over the extra dimension, we get
\be 
\label{QCDzeromodeaction}
S \, \supset \, \int d^4 x \, \frac{-1}{4  g_{\rm QCD}^2} \, G^{(0)}_{\mu \nu} \, G^{(0) \, \mu \nu} \, ,
\ee
where $G^{(0)}_{\mu \nu}$ is the field strength of the zero mode (which is identified with the standard model gluon). The tree-level contribution to its gauge coupling reads
\be 
\frac{1}{g_{\rm QCD}^2} \, = \, \frac{{\rm log}\frac{k}{\mu}}{k g_5^2} \, + \, \tau_\uv \, + \, \tau_\ir \, .
\ee
Taking also the running due to the standard model particles from the UV scale $k$ to an energy scale $Q$ into account, we get at energies below the IR scale $\mu$ (see e.g.~\cite{Csaki:2007ns})\footnote{If the cutoff is above the AdS scale, additional corrections arise from loop momenta between the two scales. These corrections can be absorbed into the parameters $g_5$, $\tau_\uv$ and $\tau_\ir$ (see e.g.~\cite{Agashe:2002bx}).}
\be 
\label{gaugecouplingrunning}
\frac{1}{g^2_{\rm QCD}(Q,\mu)} \, = \,  \frac{{\rm log}\frac{k}{\mu}}{k g_5^2}  \, - \, \frac{b_\uv}{8 \pi^2} \, {\rm log \frac{k}{Q}} \, - \, \frac{b_\ir}{8 \pi^2} \, {\rm log \frac{\mu}{Q}} \, + \, \tau_\uv \, + \, \tau_\ir \, \quad (\text{for} \, \, Q \, \lesssim \, \mu) \, .
\ee
Here $b_\uv$ is the contribution to the $\beta$-function coefficient from quarks localized near the UV brane and the gluon itself, while $b_\ir$ arises from quarks localized near the IR brane. We make the common choice that the top-bottom doublet and the right-handed top are in the IR, whereas the remaining quarks are in the UV. This then gives $b_\uv=8$ and $b_\ir=-1$.

It is straightforward to understand the origin of the terms in the above expression from the dual perspective. A gauge field in the bulk of an RS model is dual to a gauge field that weakly gauges a global symmetry of a CFT. The action in eq.~\eqref{QCDaction} then maps to
\be 
S \, \supset \, \int d^4 x \, \left(  \mathcal{L}_{\rm CFT} \, - \,  \frac{\tau_\uv }{4} \mathcal{G}_{\mu \nu} \mathcal{G}^{\mu \nu} \, + \, \mathcal{G}_\mu \mathcal{J}^\mu_{\rm CFT} \right) \, ,
\ee
where $\mathcal{L}_{\rm CFT}$ defines the CFT, $\mathcal{G}_\mu$ is the gauge field and $\mathcal{J}^\mu_{\rm CFT}$ is the current of the global symmetry.
Furthermore, the IR scale of the RS model is dual to the confinement scale of the CFT. The first term in eq.~\eqref{gaugecouplingrunning} can then be understood as arising from the CFT degrees of freedom which contribute with the $\beta$-function coefficient 
\be
\label{eq:bcft}
b_{\rm CFT} \, \equiv \,  - \frac{8\pi^2}{k g_5^2} \,. 
\ee
It depends on $\mu$ because the CFT confines at that scale and no longer contributes to the running at lower energies. 
The second term is due to the contribution from standard model particles which are fundamental and not part of the CFT sector. Instead the third term results 
from standard model particles which are (dominantly) composite states. Since the latter only arise at the confinement scale, this contribution again depends on $\mu$. Finally, the last term is due to threshold corrections at the confinement scale.

The QCD coupling thus depends on the IR scale in an RS model. Correspondingly, the scale $\Lambda_{\rm QCD}$ at which it becomes strong depends on the IR scale too. Let us define $\Lambda_{\rm QCD}$ as the scale where the QCD coupling diverges.\footnote{If we define it instead as the scale where $g_{\rm QCD}(\Lambda_{\rm QCD}(\mu), \mu)=4 \pi$, $\tau \rightarrow \tau - 1/16 \pi^2$ in eq.~\eqref{LambdatildeQCDtau} below.} From eq.~\eqref{gaugecouplingrunning}, we then find
\be 
\Lambda_{\rm QCD}(\mu) \, = \, \left( k^{b_\uv} \mu^{b_\ir} \, e^{- 8 \pi^2 \tau} \left(\frac{\mu}{k}\right)^{- b_{\rm CFT}}\right)^{1/(b_\uv + b_\ir)} \quad (\text{for} \, \, \Lambda_{\rm QCD}(\mu) \, \lesssim \, \mu) \, ,
\label{LambdatildeQCDtau}
\ee
where $\tau \equiv \tau_\uv + \tau_\ir$ and we have assumed that $\Lambda_{\rm QCD}(\mu) \, \lesssim \, \mu $. The latter condition arises because eq.~\eqref{gaugecouplingrunning} is only valid for energy scales $Q \lesssim \mu$. Indeed, the QCD scale is above the IR scale if $\Lambda_{\rm QCD}(\mu) \gtrsim \mu $ and the analysis in terms of the zero mode of the 5D gauge field is no longer justified. We discuss what happens in this regime below. 

In order to reproduce the QCD coupling today, the free parameters $g_5$ and $\tau$ need to be chosen such that $\Lambda_{\rm QCD}(\mu_{\rm min}) = \Lambda_{\rm QCD, SM}$, where $\mu_{\rm min}$ is the minimum of the radion potential and $\Lambda_{\rm QCD, SM}$ is the QCD scale today. This relation allows us to fix $\tau$ in terms of $g_5$ which then gives\footnote{Note that $\tau$ is always positive for the parameters that we consider. }
\be 
\Lambda_{\rm QCD}(\mu) \, = \,\Lambda_{\rm QCD, SM} \, \left(\frac{\mu}{\mu_{\rm min}}\right)^n \qquad (\text{for} \, \,  \Lambda_{\rm QCD}(\mu)\, \lesssim \, \mu) \, ,
\label{LambdatildeQCD}
\ee
where 
\be 
\label{eq:nversusbeta}
n \, \equiv \, \frac{b_{\ir} -b_{\rm CFT}}{b_\uv + b_\ir} \, .
\ee
The size of $g_5$ and thus $b_{\rm CFT}$ and $n$ is limited by the requirement that the KK decomposition is sensible in the effective 4D theory. Indeed, since the gauge coupling in 5D is an irrelevant operator, the theory is expected to become strongly coupled at the scale $\Lambda_c \sim 16 \pi^2 /g_5^2$. Demanding that at least one KK mode is still in the perturbative regime, $\pi k \lesssim \Lambda_c$, we find the condition $g_5^2 k \lesssim 16 \pi$. This translates to $n \gtrsim 0.1$.

We will be interested in the case $n<1$. For the radion at the minimum of its potential, $\mu = \mu_{\rm min}$, the QCD scale is given by  $\Lambda_{\rm QCD}(\mu_{\rm min})=\Lambda_{\rm QCD, SM} \ll \mu_{\rm min}$. Then moving the radion away from the minimum to smaller values, the QCD scale decreases. 
For $n<1$, $\Lambda_{\rm QCD}(\mu)$ decreases slower than linearly with decreasing $\mu$ though and eventually both become comparable. For even smaller radion values, the condition for eq.~\eqref{LambdatildeQCD} is then no longer satisfied. In order to see what happens for $\Lambda_{\rm QCD} \gtrsim \mu$, it is again useful to consider the dual perspective. The dependence of the QCD scale on $\mu$ arises in this description, because the CFT confines at the scale $\mu$ and (most of) its states no longer contribute to the running  
of the QCD coupling at energies below $\mu$. In addition, such a dependence also results because some states (corresponding to the IR-localized quarks) only arise at the scale $\mu$ and then contribute to the running at lower energies. Since for $\Lambda_{\rm QCD} > \mu$, QCD confines at higher energies than the CFT, in this regime $\Lambda_{\rm QCD}$ will become independent of $\mu$.\footnote{
In the 5D description, eq.~\eqref{gaugecouplingrunning} is the running gauge coupling of the zero-mode of the 5D gauge field. For energies above the KK mass scale, such a coupling is ill-defined. Instead one can define the coupling in this regime via the gauge field correlator with endpoints restricted to the UV brane \cite{Goldberger:2003mi,Goldberger:2002cz,Goldberger:2002hb}. One then in particular finds that the loop corrections become independent of the IR scale (or KK mass scale) for energies above that scale (see e.g.~sec.~III B in \cite{Goldberger:2002hb}). 
} By continuity, we then expect
\be 
\label{LambdatildeQCD2}
\Lambda_{\rm QCD}(\mu) \, = \, \Lambda_{\rm QCD}(\mu_c) \qquad (\text{for} \, \,\Lambda_{\rm QCD}(\mu) \, \gtrsim \, \mu) \, .
\ee
Here $\mu_c$ is the IR scale for which $\Lambda_{\rm QCD}(\mu_c)$ is so large that eq.~\eqref{LambdatildeQCD} is no longer applicable. We parametrise our ignorance where precisely this happens by a parameter $n_c$ and define $\mu_c$ as the IR scale for which
\be 
\label{muccondition}
\Lambda_{\rm QCD}(\mu_c) \, = \, n_c \, \mu_c \, . 
\ee
The conditions of validity for eqs.~\eqref{LambdatildeQCD} and \eqref{LambdatildeQCD2} then become $\Lambda_{\rm QCD}(\mu) \lessgtr n_c \, \mu$. For $n<1$, this is equivalent to $\mu \gtrless \mu_c$. Using eq.~\eqref{LambdatildeQCD}, we find
\be
\label{muc}
\mu_c \, = \, \mu_{\rm min} \, \left(\frac{\Lambda_{\rm QCD, SM}}{n_c\mu_{\rm min}}\right)^{\frac{1}{1-n}}  \, .
\ee
We expect that $n_c$ is larger than 1. Indeed, the description in terms of the zero mode of the 5D gauge field should not break down immediately when the QCD confinement scale becomes larger than the IR scale, $\Lambda_{\rm QCD}(\mu)> \mu$. We instead expect that this description becomes no longer applicable only once the QCD confinement scale reaches the mass scale of the first KK mode of the 5D gauge field, $\Lambda_{\rm QCD}(\mu) \gtrsim m_{\rm KK}$. This would imply $n_c \sim \pi$.

\begin{figure}[t]
\centering
\includegraphics[width=12cm]{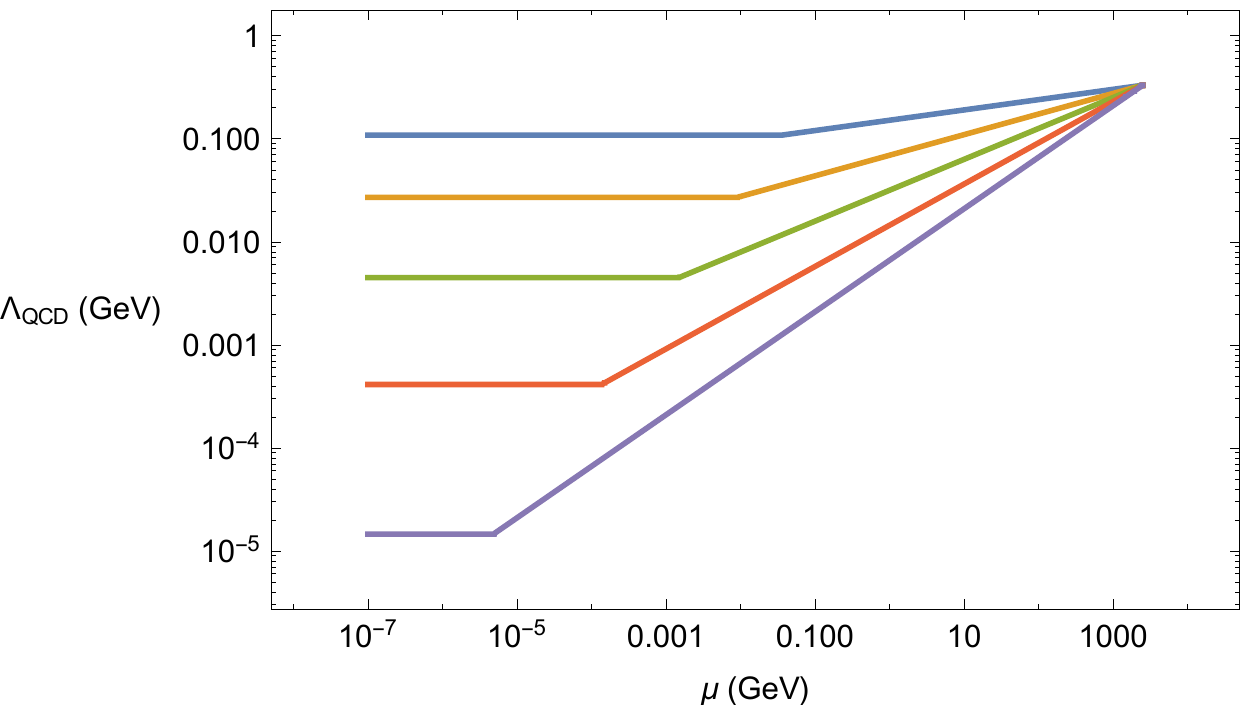}
\caption{\label{fig:LambdatildeQCD} \small Schematic plot of the QCD confinement scale $\Lambda_{\rm QCD}$ as a function of the IR scale $\mu$ for $\mu_{\rm min}=2.5 \,$TeV, $n_c=3$ and $n=0.1,0.2,0.3,0.4,0.5$ (in blue, yellow, green, red, purple).}
\end{figure}

We plot $\Lambda_{\rm QCD}(\mu)$ as determined in this section for $\mu_{\rm min}=2.5 \,$TeV, $n_c=3$ and different values of $n$ in fig.~\ref{fig:LambdatildeQCD}. Starting from $\mu=\mu_{\rm min}$, it initially decreases with decreasing $\mu$ according to eq.~\eqref{LambdatildeQCD}. It then eventually reaches the value in eq.~\eqref{muccondition}, after which it stays constant. We expect that the change between the scalings in eqs.~\eqref{LambdatildeQCD} and \eqref{LambdatildeQCD2} will be smoother than shown in the plot. 
 
\section{Contribution of the QCD condensates to the radion potential}
\label{sec:QCDcont}
As we will discuss in more detail in sec.~\ref{sec:PT}, if the radion potential is solely determined by the Goldberger-Wise field, the phase transition in RS models can only complete in small regions of parameter space. For most choices of parameters the radion instead remains stuck in the wrong vacuum and the universe enters an inflationary phase. This lowers the temperature of the surrounding plasma. Eventually the temperature reaches the QCD scale and QCD confines. 
As is well-known, this generates condensates for the gluon and the light quarks. We now discuss how these can affect the radion potential. 

The gluon condensate was determined in ref.~\cite{Narison:2011xe} as (see also \cite{Narison:2010cg,Narison:2011rn})\footnote{
In the relevant literature, typically a convention is used where the gauge coupling appears in the covariant derivative. Then values for the expectation value $\langle \alpha_s G_{\mu \nu} \, G^{\mu \nu} \rangle$ are quoted. In our convention (cf.~eq.~\eqref{QCDzeromodeaction}), this leads to the factor $4 \pi$ in eq.~\eqref{GluonCondensate}.
} 
\be
\label{GluonCondensate}
\langle  G^{(0)}_{\mu \nu} \, G^{(0) \, \mu \nu} \rangle \, = \, 4 \pi \, \cdot \,(7  \pm  1) \cdot  10^{-2} \, \, \text{GeV}^4 \, ,
\ee
where the index $(0)$ denotes that the gluon is the zero-mode of a KK tower in our 5D model. 
A somewhat smaller value was given in ref.~\cite{Ioffe:2005ym}, though with a significantly larger error, while lattice studies in refs.~\cite{DiGiacomo:1981lcx,Campostrini:1989uj,Rakow:2005yn} find a range of values. We use the result in eq.~\eqref{GluonCondensate} for definiteness in the following but our analysis is not very sensitive to $\mathcal{O}(1)$-variations in the gluon condensate. 
The condensates of the light quarks, on the other hand, are found to be \cite{Ioffe:2005ym}
\be
\label{quarkcondensate}
 \langle  \,\overline{\psi}^{(0)}_{u,d} \, \psi^{(0)}_{u,d} \, \rangle \, = \, - (1.65  \pm 0.15) \cdot  10^{-2} \, \, \text{GeV}^3 \, .
\ee
The condensate of the strange quark is smaller by about a factor $0.8$  \cite{Ioffe:2005ym}. These condensates contribute to the trace of the energy-momentum tensor
\be 
\label{TraceEnergyMomentumTensor}
T^\rho_{\, \rho} \, \supset \, -\frac{b_{\rm QCD}}{32 \pi^2}\,  G^{(0)}_{\mu \nu} \, G^{(0) \, \mu \nu}  \, + \,  \sum_{\rm quarks} m_q \,  \overline{\psi}_i^{(0)} \psi_i^{(0)} \, .
\ee
The first term is due to the scale anomaly of QCD, where $b_{\rm QCD}$ is the $\beta$-function coefficient of QCD, and the sum in the second term is over all quarks that form a condensate, where $m_q$ denotes their masses. The trace of the energy-momentum tensor in turn relates to the energy density as 
\be
\label{EnergyDensity}
V \,  = \, \frac{1}{4} \, \langle \, T^\rho_{\, \rho} \, \rangle \, .
\ee
Once QCD confines, it thus contributes to the energy density of the universe. Since in the RS model the scale at which QCD becomes strongly coupled depends on the radion, the size of the condensates and thus their contribution to the energy density depends on it too. On dimensional grounds, we expect that 
\begin{align}
\label{GluonCondensateScaling}
& \langle  G^{(0)}_{\mu \nu} \, G^{(0) \, \mu \nu} \rangle \, \sim \, \left(\Lambda_{\rm QCD}(\mu)\right)^4 \, , \\
& \langle  \,\overline{\psi}^{(0)}_{u,d} \, \psi^{(0)}_{u,d} \, \rangle \, \sim \, \left(\Lambda_{\rm QCD}(\mu)\right)^3 \, .
\end{align}
Following from eqs.~\eqref{TraceEnergyMomentumTensor} and \eqref{EnergyDensity}, this leads to an additional contribution from QCD to the radion potential.

We are thus interested in situations where the phase transition of RS models happens at temperatures at or below the QCD scale. Electroweak symmetry is then generically broken simultaneously. Correspondingly, we in principle need to analyze the phase transition in the two-field potential for the radion $\mu$ and the Higgs $\langle H \rangle$. The Higgs then in particular affects the potential from the quark condensates via the quark masses. Let us for the moment assume that the radion tunnels into its minimum first and the Higgs only follows afterwards. Then $\langle H \rangle =0$ during the phase transition of the radion and the contribution from the quark condensates vanishes. 
For the contribution from the gluon condensate, we can estimate the prefactor in eq.~\eqref{GluonCondensateScaling} by matching with eq.~\eqref{GluonCondensate} for $\Lambda_{\rm QCD} = \Lambda_{\rm QCD, SM} \simeq 330 \, \text{MeV}$ \cite{Patrignani:2016xqp}. 
We then find
\be 
V_{\rm QCD}(\mu,\langle H \rangle=0) \, \approx \, -  \, \frac{b_{\rm QCD}}{17} \, \left(\Lambda_{\rm QCD}(\mu) \right)^4 
\label{NewContributionRadionPotential}
\ee
for the contribution of the gluon condensate to the radion potential. Several comments are in order: The prefactor in this relation could have an additional dependence on $\Lambda_{\rm QCD}$ and thus $\mu$. However, we expect that the resulting change with $\mu$ in the prefactor is at most of order 1. We will later see that our results are relatively insensitive to changes of this (or even somewhat larger) magnitude. More important is that the gluon condensate $ \langle  G^{(0)}_{\mu \nu} \, G^{(0) \, \mu \nu} \rangle$ remains positive for all confinement scales, so that the prefactor does not change sign. But since this quantity should (at least in principle) be calculable using a path integral in Euclidean space-time, this is trivially satisfied (see the discussion in sec.~6.9 in ref.~\cite{Shifman:1978bx}).
The positivity of the gluon condensate also makes intuitive sense because it means that the energy density is lowered during confinement (the quark condensates give a comparatively smaller contribution).

Note that all quarks are massless along the direction $\langle H \rangle=0$. The relevant $\beta$-function coefficient in eq.~\eqref{NewContributionRadionPotential} therefore is $b_{\rm QCD}=7$.
Correspondingly instead of $\Lambda_{\rm QCD, SM}$, which is the scale where the QCD coupling diverges if 3 flavours are light at that scale (the other 3 flavours are decoupled at their respective masses), we need to use $\Lambda_{\rm QCD, 0} \simeq 90 \, \text{MeV}$ for the case of 6 light flavours \cite{Patrignani:2016xqp} in eqs.~\eqref{LambdatildeQCD} and \eqref{LambdatildeQCD2}. This gives
\be 
V_{\rm QCD}(\mu,\langle H \rangle=0) \, \approx \, -  \, \frac{b_{\rm QCD}}{17} \, \cdot \, 
\begin{cases}
 \Lambda_{\rm QCD, 0}^4  \, \left(\frac{\mu}{\mu_{\rm min}} \right)^{4n}  &\quad \text{for $\mu > \mu_c$} \\
 \; \left(\Lambda_{\rm QCD} (\mu_c)\right)^4  &\quad \text{for $\mu < \mu_c$} \, .
\end{cases}
\label{NewContributionRadionPotential2}
\ee

\begin{figure}[t]
\centering
\includegraphics[width=7.5cm]{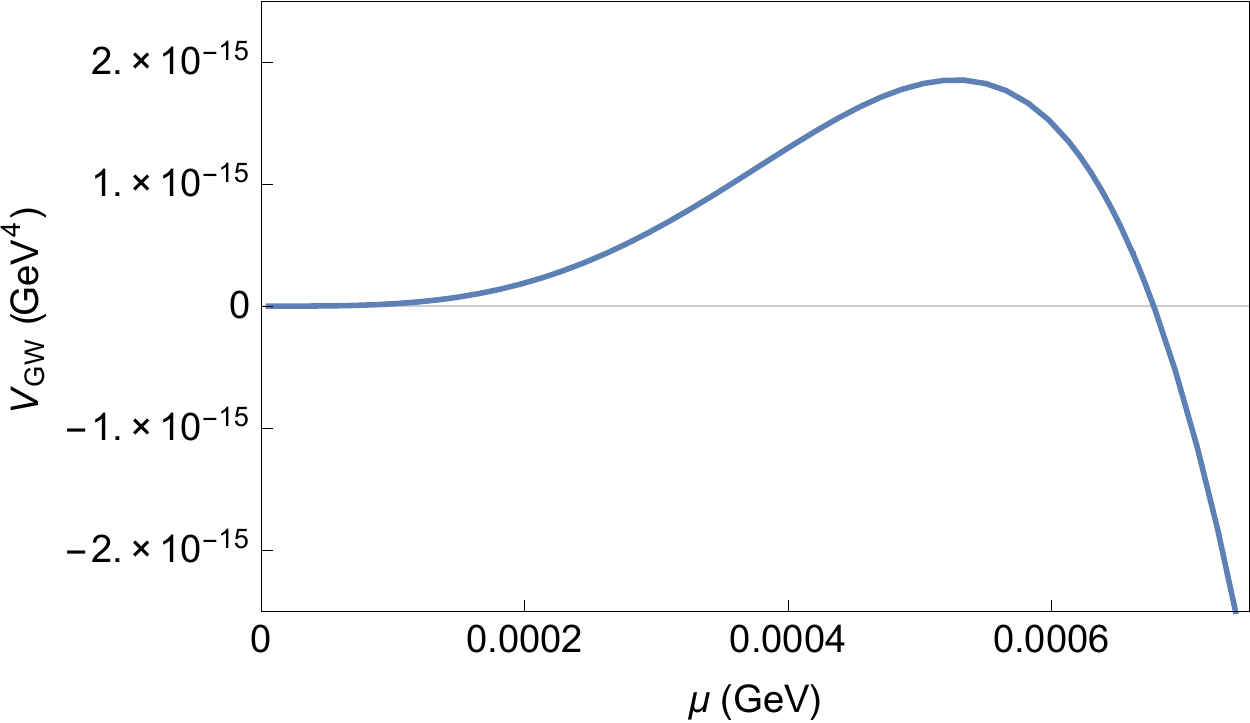}
\hspace*{.4cm}
\includegraphics[width=7.5cm]{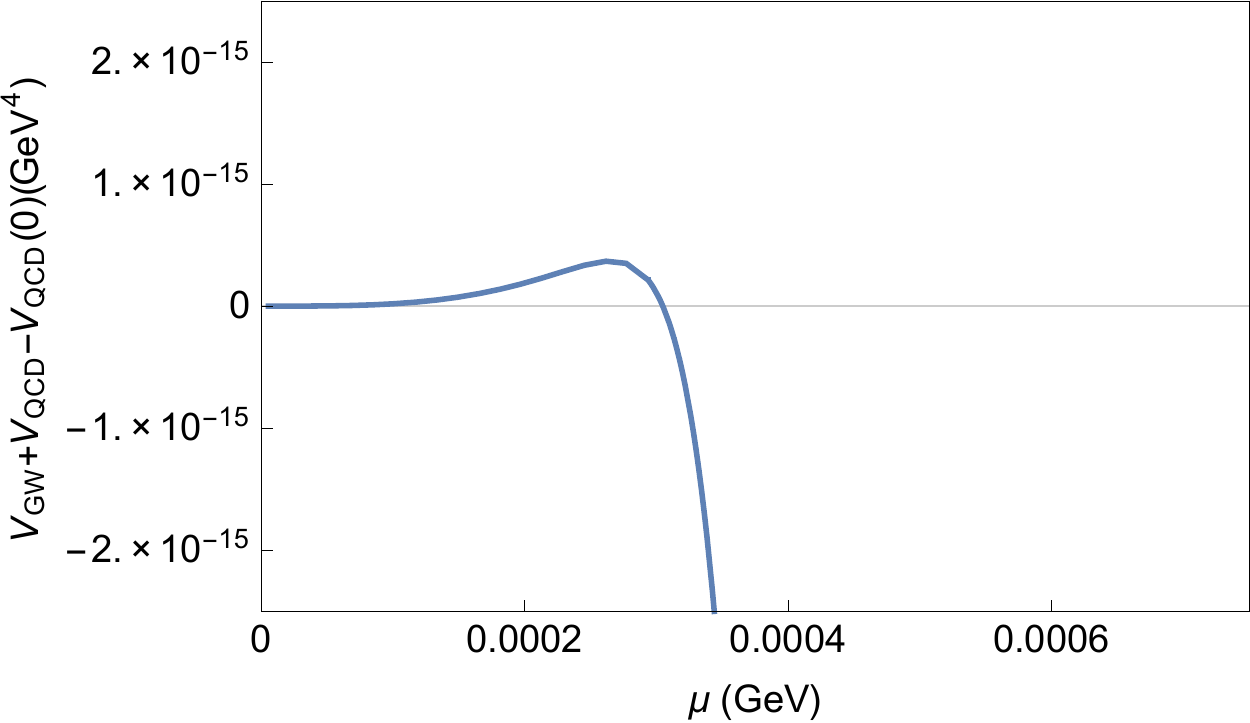}
\caption{\label{fig:potential} \small The radion potential plotted around the Goldberger-Wise barrier without (left) and with (right) the contribution from the gluon condensate for $\mu_{\rm min}=2.5\,$TeV, $n = 0.15 $, $n_c=3$, $\epsilon = 1/20$, $v_{\rm IR}=1$ and $\delta=-1/2$. The combined potential is negative near the origin because the gluon condensate contributes with a negative sign, while the Goldberger-Wise potential vanishes there. For better comparison, we have shifted the combined potential to make it vanish at the origin too. 
Notice that the barrier does not completely disappear even with the contribution from the gluon condensate.}
\end{figure}

The energy density in the minimum of the Goldberger-Wise potential is given by 
\be
\label{VGWmumin}
V_{\rm GW}(\mu_{\rm min}) \, \simeq \, - \epsilon \, v_\ir^2 \, \mu_{\rm min}^4 \, \left(\sqrt{\epsilon + \frac{\epsilon^2}{4} - \frac{\delta}{v_\ir^2}} \, - \, \frac{\delta}{2  v_\ir^2}\right) \, .
\ee
This is typically much bigger than $V_{\rm QCD}(\mu_{\rm min})$ since $\mu_{\rm min} \gg \Lambda_{\rm QCD,SM}$. The new contribution from QCD is thus negligible near the minimum of the radion potential. However, the Goldberger-Wise potential goes approximately like $\mu^4$, while the potential from the gluon condensate is proportional to $ \mu^{4n}$. For $n<1$, the importance of the latter relative to the former thus grows with decreasing $\mu$. 
Since the gluon condensate contributes with a negative sign to the energy density, it can then partly remove the barrier in the Goldberger-Wise potential between the origin and the minimum. This is borne out in fig.~\ref{fig:potential}, where we plot the radion potential without and with the contribution from the gluon condensate near the Goldberger-Wise barrier for $n = 0.15 $ (and $n_c=3$, $\mu_{\rm min}=2.5\,$TeV, $v_{\rm IR}=1, \epsilon = 1/20$). The gluon condensate indeed removes a significant part of the barrier and, more generally, changes the shape of the potential. 
Notice that it does not remove the barrier completely though and a small barrier remains. The reason is that the gluon condensate becomes independent of $\mu$ for $\mu \lesssim \mu_c$ as discussed previously, while the Goldberger-Wise potential grows approximately like $\mu^4$ near the origin. 
Since a barrier remains, the phase transition is still first order. We then need to calculate the bounce in order to see if the tunneling rate is sufficiently high for the phase transition to complete in the early universe. But from the plot of the potential, we expect that the QCD contribution can significantly increase the tunneling rate. We will see later that this is indeed the case.

Let us next discuss the case where both $\mu$ and $\langle H \rangle$ change simultaneously during the phase transition. The contribution from the gluon condensate is then still given by eq.~\eqref{NewContributionRadionPotential}. But the $\beta$-function coefficient becomes a (stepwise) function of $\mu$ and $\langle H\rangle$ since it depends on the number of light fermions near the QCD scale $\Lambda_{\rm QCD}$. In addition, the quark condensates now contribute to the potential. Matching with eq.~\eqref{quarkcondensate} for $\Lambda_{\rm QCD}=\Lambda_{\rm QCD, SM}$ gives the estimate
\be 
V_{\rm QCD}(\mu,\langle H \rangle) \, \supset \,  - \, \frac{1}{2} \,\sum_{\rm quarks} \, y_q \, \langle H \rangle \left(\Lambda_{\rm QCD}(\mu)\right)^3  \, .
\label{NewContributionRadionPotential3}
\ee
As we discuss below, this relation is a priori only valid for $\Lambda_{\rm QCD}(\mu) \lesssim \mu$. Also again we expect some additional $\mathcal{O}(1)$-dependence on $\mu$ in this relation for $\Lambda_{\rm QCD}$ different from $\Lambda_{\rm QCD, SM}$. The sum is over all quarks with $ m_q = y_q \langle H \rangle \lesssim \Lambda_{\rm QCD}(\mu)$. Near the minimum of the combined radion-Higgs potential at $\mu = \mu_{\rm min}$ and $\langle H \rangle=v_{\rm EW}$, this sum is dominated by the strange quark with $y_q \approx 10^{-3}$. On the other hand, in the region of the potential where $\langle H \rangle \lesssim \Lambda_{\rm QCD}$, even the top quark condenses and contributes. In order to compare with the contribution from the gluon condensate, let us consider two sample trajectories near the minimum of the two-field potential. For   
\be 
\label{trajectory1}
\langle H \rangle \, = \, v_{\rm EW} \, \left(\frac{\mu}{\mu_{\rm min}}\right)^n \, ,
\ee
the ratio $\langle H \rangle / \Lambda_{\rm QCD}(\mu)$ remains constant and the strange quark condensate dominates over the other quark condensates everywhere along the trajectory. We then see from eqs.~\eqref{NewContributionRadionPotential} and \eqref{NewContributionRadionPotential2} that the gluon and quark condensates contribute approximately equally to the potential. Let us next consider the trajectory
\be 
\label{trajectory2}
\langle H \rangle \, = \, v_{\rm EW} \, \frac{\mu}{\mu_{\rm min}}
\ee
along the minimum of the Higgs potential.\footnote{We note that this trajectory is along the minimum of the Higgs potential only if the Higgs mass parameter $m_H^2$ is independent of $\mu$. However, various phenomenological constraints require that this mass is much smaller than its natural value, $|m_H^2| \ll e^{-2k y_\ir}M_5^2$ (or a similar cutoff). In absence of a dynamical mechanism to generate this (little) hierarchy, one needs an accidental cancellation among different contributions to $m_H^2$ to bring it down to the required value. It is then expected that this accidental cancellation only happens for $\mu$ close to $\mu_{\rm min}$ and that the Higgs mass parameter is brought back to its natural value for different $\mu$. This would change the trajectory along the minimum of the Higgs potential to $\langle H \rangle \sim \mathcal{O}(1)\cdot \mu$. We thank Jay Hubisz for emphasizing this to us.\label{HiggsMassTermFootnote}} 
The ratio $\langle H \rangle / \Lambda_{\rm QCD}(\mu)$ then decreases for $n <1$ when going along this trajectory from $\mu=\mu_{\rm min}$ towards $\mu = 0$ and more and more quark flavours condense. This increases the importance of the quark condensates for the potential relative to the gluon condensate. On the other hand, this is counteracted by the fact that eq.~\eqref{NewContributionRadionPotential2} now decreases proportional to $\mu^{1+3n}$ with decreasing $\mu$, while eq.~\eqref{NewContributionRadionPotential} still scales as $\mu^{4n}$. 

Nevertheless, it is possible that there are regions of parameter space and trajectories in the two-field potential for which the quark condensates dominate over the gluon condensate. However, we will refrain from analyzing this quantitatively. As earlier in this section, we will instead focus on the tunneling path along the $\smash{(\mu,\langle H \rangle=0)}$-direction for the radion and assume that the Higgs only later obtains a VEV. We can then restrict ourselves to the gluon condensate. Let us assume that, for a given point in parameter space, the tunneling action along this direction is sufficiently small to allow the phase transition to complete. If the actual tunneling path in the two-field potential differs from this direction, it necessarily has a smaller tunneling action and therefore provides a successful phase transition too. 
Focusing on the path along the $\smash{(\mu,\langle H \rangle=0)}$-direction and the gluon condensate is therefore sufficient for showing that QCD can significantly enlarge the regions of parameter space where the RS phase transition completes in the early universe. Other paths and the quark condensates can only open up more parameter space. 

In addition, there are technical reasons for focusing on the gluon condensate: 
It is in particular less clear that the sign of the quark condensate does not change when $\Lambda_{\rm QCD}$ becomes different from $\Lambda_{\rm QCD, SM}$ (contrary to the case for the gluon condensate). But only for a negative sign as in eq.~\eqref{quarkcondensate} can the resulting contribution to the radion potential remove the barrier and thereby help with the phase transition. Furthermore, the derivation of eqs.~\eqref{TraceEnergyMomentumTensor} and \eqref{NewContributionRadionPotential2} assumes that we can perform a KK expansion of the 5D fields and then add confinement as a small perturbation. If the QCD scale becomes larger than the KK scale, this assumption is no longer justified. 
In particular, higher-dimensional operators involving colored KK modes then grow with $\Lambda_{\rm QCD}/\mu$ and this description is thus no longer under control. It is unclear how the Higgs couples to the quark condensates in this regime.

\section{The phase transition in Randall-Sundrum models}
\label{sec:PT}

We next review the phase transition that happens in RS models when they cool from temperatures above to temperatures below the IR scale. In this section, we ignore the effect of the gluon condensate on the radion potential and the phase transition.  Readers familiar with this may jump straight to the next section, where we include the QCD effect and which presents our main results. 

At temperatures far above the IR scale $\mu_{\rm min}$, the geometry of the Randall-Sundrum model is deformed into AdS-Schwarzschild. This space has a black hole horizon instead of the IR brane. The position of this horizon, or equivalently its Hawking temperature $T_H$, is the relevant field variable in the AdS-Schwarzschild phase (similar to the radion $\mu$ in the Randall-Sundrum phase). Its potential is given by the free energy of AdS-Schwarzschild \cite{Creminelli:2001th}\footnote{Bulk fields like the Goldberger-Wise scalar give additional, smaller contributions (see ref.~\cite{Creminelli:2001th}).}
\be 
\label{AdS-S-Potential}
V_{\rm AdS-S}(T_H) \, = \, \frac{3}{8} \pi^2 N^2  T_H^4 \, - \, \frac{1}{2} \pi^2 N^2 T_H^3 T \, ,
\ee
where $T$ denotes the ambient temperature. As expected, this potential is minimized for $T_H=T$. Notice that the energy in this minimum increases with decreasing temperature. Eventually the temperature in the early universe has cooled so much that the minimum becomes shallower than the minimum of the Goldberger-Wise potential. Subsequently a phase transition from AdS-Schwarzschild to the Randall-Sundrum space can take place. This becomes energetically possible at the critical temperature
\be 
\label{CriticalTemperature}
T_c \, = \, \left(\frac{-8 \, V_{\rm GW}(\mu_{\rm min})}{\pi^2 N^2}\right)^{1/4} \, ,
\ee
where the energy density $V_{\rm GW}(\mu_{\rm min})$ in the minimum of the Goldberger-Wise potential is given in eq.~\eqref{VGWmumin}.

\begin{figure}[t]
\centering
\includegraphics[width=10cm]{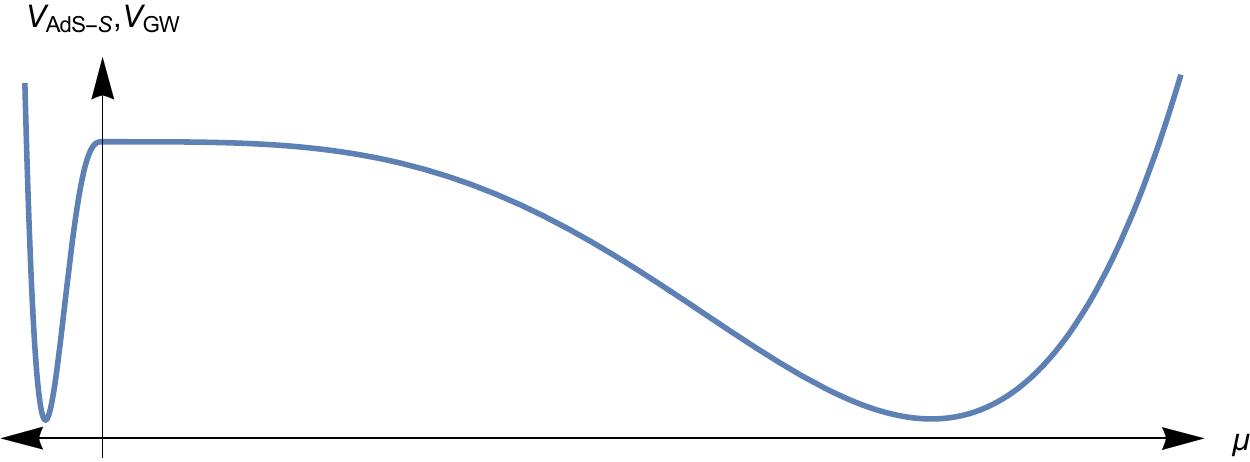}
\caption{\label{fig:CombinedPotential} \small Schematic plot of the combined potential for the AdS-Schwarzschild and Randall-Sundrum spaces, glued together at the origin and evaluated at the critical temperature. }
\end{figure}

The two spaces have different topologies since AdS-Schwarzschild is simply connected whereas the Randall-Sundrum space is not. But they can be smoothly connected by sending respectively the horizon and the IR brane to infinity, $T_H,\mu\rightarrow 0$, which in both cases gives pure AdS$_5$ (cut off by a brane in the UV). As argued in ref.~\cite{Creminelli:2001th}, it is then plausible to expect that the dominant bounce which mediates the phase transition interpolates between the two spaces via pure AdS$_5$. The potential which governs this bounce is obtained by gluing the potentials for $T_h$ and $\mu$, $V_{\rm AdS-S}$ and $V_{\rm GW}$, together at the origin, $T_H=\mu=0$.\footnote{Note that in the region $0 \leq \mu \lesssim T$, temperature corrections to the radion potential are not under control as the effective 4D description breaks down. However, we are mainly interested in the potential at the nucleation temperature $T_n$ which is much smaller than $\mu_{\rm min}$ (see below). The  region which is not under control is therefore small and can be neglected in calculating the bounce.} In fig.~\ref{fig:CombinedPotential}, we plot this combined potential at the critical temperature. Notice that the AdS-Schwarzschild phase leads to a barrier that separates the two minima in the potential. This comes in addition to the barrier in the Goldberger-Wise potential which we have mentioned before and which is already present at zero temperature. 
The phase transition therefore is first-order and proceeds via the nucleation of bubbles. The rate with which these bubbles form is given by 
\be 
\label{BubbleNucleationRate}
\Gamma_n \, = \, \Gamma_0 \, e^{-\mathcal{S}} \, ,
\ee
where $\Gamma_0$ is of order of  the fourth power of the relevant energy scale of the potential and $\mathcal{S}$ is the bubble action. For $O(4)$-symmetric bubbles, $\Gamma_0$ is a function of the characteristic bubble size, the field value $\mu_{\rm r}$ to which the field tunnels 
and the second derivative of the potential along the tunneling path \cite{Coleman:1977py,Callan:1977pt,Linde:1981zj}. The scale of all these quantities is set by $\mu_{\rm r}$. For $O(3)$-symmetric bubbles, an additional dimensionful quantity is the ambient temperature \cite{Linde:1981zj}. Since these bubbles are only important for relatively high temperatures, this is typically again of order $\mu_{\rm r}$. In addition, $\Gamma_0$ can depend on the dimensionless quantities $N,\epsilon, v_\ir, \delta$. Since $\Gamma_0$ only enters logarithmically into the relation for the required bubble action $\mathcal{S}$ (see below), though, these parameters can be neglected.

The phase transition can only complete if the rate of bubble nucleation $\Gamma_n$ becomes larger than the Hubble rate per horizon time and volume $H^4$. This leads to the condition on the bubble action 
\be 
\label{PhaseTransitionBubbleAction}
\mathcal{S} \, \lesssim \, 4 \, \log \left(\frac{\mu_{\rm r} M_{\rm Pl}}{ \, \mu^2_{\rm min}}\right) \, ,
\ee
where we have used the estimate $\Gamma_0 \sim \mu_r^4$ and that $H \sim  \mu_{\rm min}^2/M_{\rm Pl}$ during the phase transition. Note that the latter relation applies for both temperatures near and far below the critical temperature. The Hubble rate in the former case is driven by the AdS-Schwarzschild phase, while in the latter case it is determined by the cosmological constant which arises from the universe being stuck in the wrong vacuum (cf.~eqs.~\eqref{VGWmumin} and \eqref{AdS-S-Potential}). If the radion potential is of the Goldberger-Wise type, the radion typically tunnels to near the minimum of the potential so that $\mu_r \sim \mu_{\rm min}$. This leads to the criterion $\mathcal{S} \lesssim 140$. As we will discuss in sec.~\ref{sec:QCDeffect}, if the gluon condensate modifies the potential near the origin, the radion instead tunnels to the much smaller value $\mu_r \sim \Lambda_{\rm QCD}(\mu_c)$. The resulting criterion $\mathcal{S} \lesssim 50 - 90$ for $n=0.1 - 0.5$ is more stringent. The modified potential from the gluon condensate will have a much smaller tunneling action, though, which can then still satisfy the tightened criterion for the phase transition to complete.

One can distinguish two types of bubbles that form during the phase transition.  
If the spatial size of the bubble is much larger than the radius $T^{-1}$ of the time direction, it has $O(3)$ symmetry. The bubble action is then given by $\mathcal{S} = S_3/T$, where $S_3$ is the action of the spatial part. In the opposite case, the bubble has $O(4)$ symmetry and its action reads $\mathcal{S}=S_4$. 
The action for both bubble types generically depends on the temperature and thus changes as the universe expands and cools. For $O(4)$-symmetric bubbles, it decreases with the temperature as the energy difference between the false and true minima then grows (cf.~eqs.~\eqref{VGWmumin} and \eqref{AdS-S-Potential}). This effect is partly counteracted for $O(3)$-symmetric bubbles due to the explicit $T^{-1}$-suppression of the action. The latter is therefore typically minimized for temperatures not too far below the critical temperature.

Let us first consider bubbles with $O(4)$ symmetry. If we want to determine whether the phase transition can complete, it is sufficient to calculate their action at zero temperature since this minimizes the action. The AdS-Schwarzschild part of the instanton vanishes in this limit and we can use the origin of the Goldberger-Wise potential as the initial state corresponding to the false vacuum. The radial profile $\mu(r)$ of the bubbles is then obtained by solving the bounce equation
\be 
\label{BounceEquation}
\frac{3 N^2}{2 \pi^2} \left(\frac{d^2 \mu}{dr^2} \, +\, \frac{3}{r} \, \frac{d \mu}{dr}\right) \, = \, \frac{d V_{\rm GW}}{d \mu} \, ,
\ee
where $r = \sqrt{\vec{x}^2 +t^2}$ is the radial distance from the center of the bubble and the boundary conditions are $\mu(r)\rightarrow 0$ for $r\rightarrow \infty$ and $d\mu/dr=0$ at $\mu=0$. The bubble action follows from the integral
\be 
\label{S4Integral}
S_4 \, = \, 2 \pi^2 \int r^3 dr \left[\frac{3 N^2}{4 \pi^2}\left(\frac{d \mu}{dr}\right)^2 \, + \, V_{\rm GW}(\mu) \right] \, .
\ee

\begin{figure}[t]
\centering
\includegraphics[width=7.8cm]{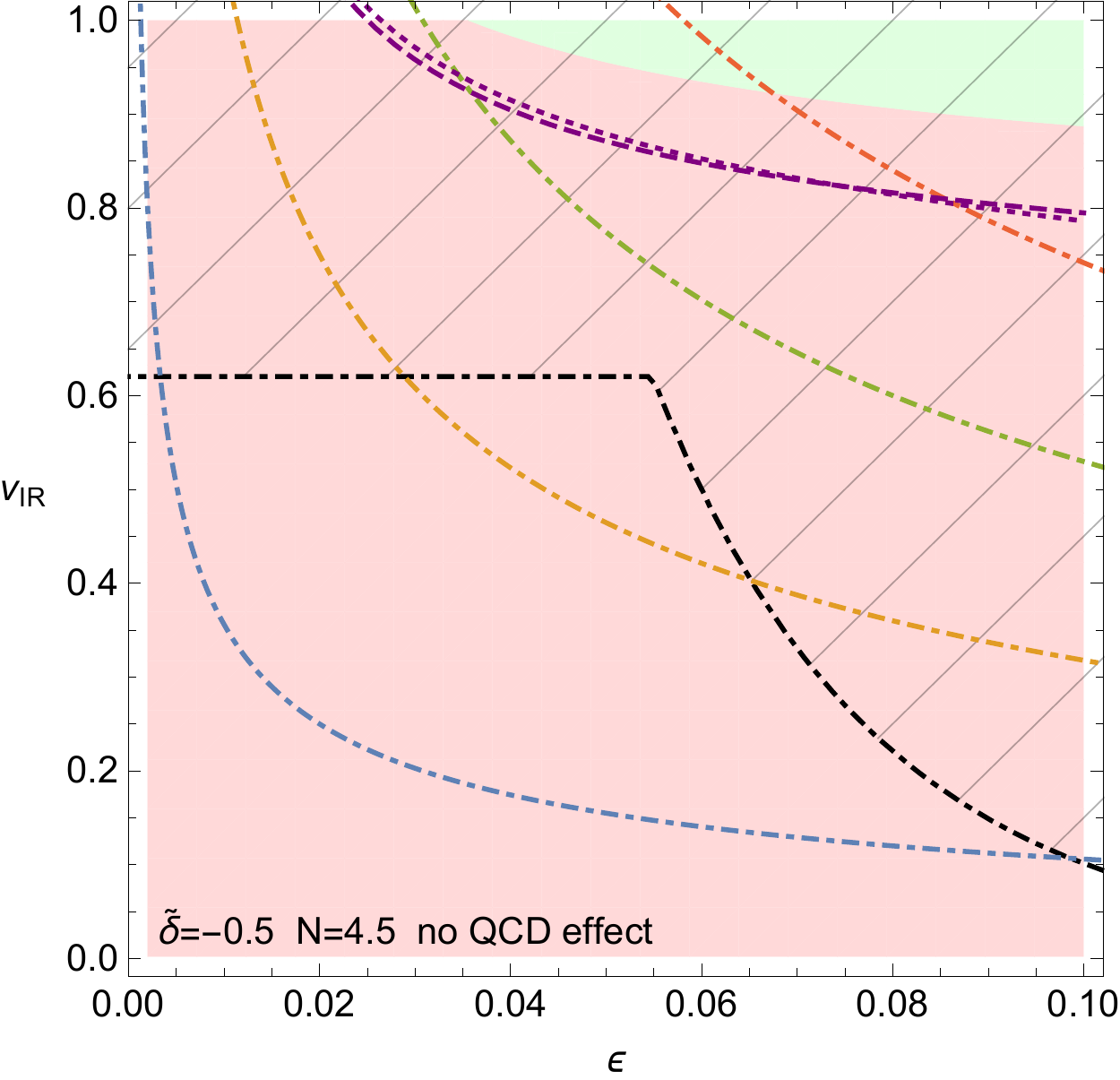}
\hspace{.3cm}
\includegraphics[width=7.8cm]{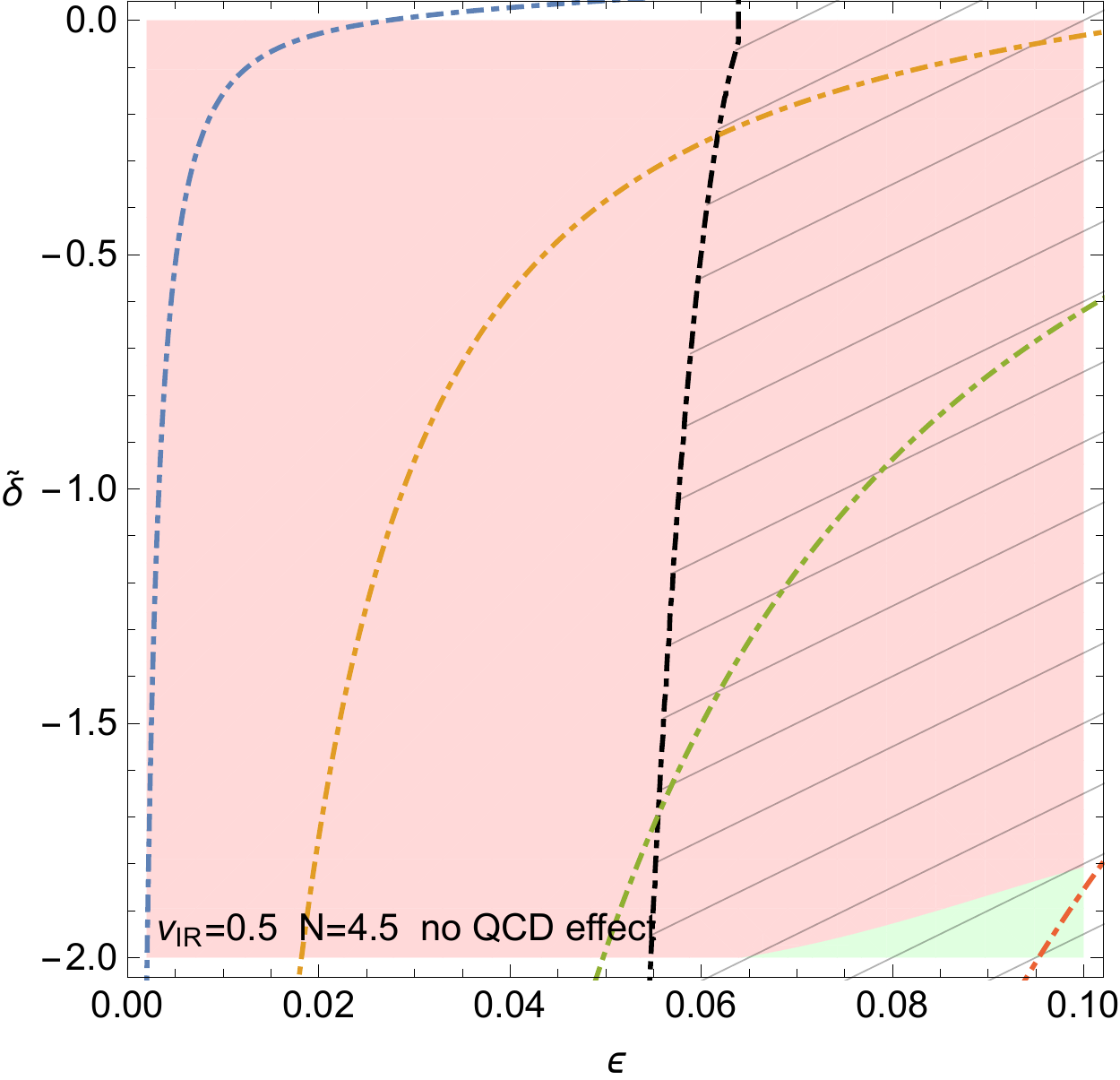}
\caption{\label{fig:Resultsna0}  \small Results for $\mu_{\rm min}=2.5\,$TeV and $N=4.5$ without the QCD effect. For the left panel, we have fixed $\tilde{\delta}=-0.5$, and for the right panel, $v_\ir =0.5$. Regions where the phase transition can complete via the nucleation of $O(4)$-symmetric bubbles are shown in green, while regions where the nucleation rate is too low are colored in red. Regions above the purple dashed and dotted lines are allowed according to the analytical estimate of the bubble action for $O(4)$- and $O(3)$-symmetric bubbles, respectively. In the hashed region (above the black, dashed line in the left panel and to right of the black, dashed line in the right panel), the backreaction constraint is not fulfilled. The blue, orange, green, red dashed-dotted lines (from bottom to top in the left panel and in the reversed order in the right panel) correspond to the radion mass being $m_{\rm radion} =  200\, \text{GeV}, 600\, \text{GeV}, 1\, \text{TeV}, 1.4\, \text{TeV}$, respectively.}
\end{figure}

We have numerically calculated the resulting action for $\mu_{\rm min}=2.5\,$TeV, $N=4.5$ and different values of $\epsilon$, $v_\ir$ and $\delta$ (ignoring the QCD effect). In order to stay in the window for $\delta$ above eq.~\eqref{GWminmax} for varying $v_\ir$, we choose the parametrization $\delta = \tilde{\delta} \, v_\ir^2$. In the left panel of fig.~\ref{fig:Resultsna0}, we then fix $\tilde{\delta} = -0.5$ and show results in the $(v_\ir-\epsilon)$-plane. In the right panel, $v_\ir = 0.5$ and results are plotted in the $(\tilde{\delta}-\epsilon)$-plane.
Regions where the action satisfies the criterion in eq.~\eqref{PhaseTransitionBubbleAction} and where the phase transition can thus complete are shown in green. In the remaining parameter space, shown in red, the phase transition does not complete and the radion instead remains stuck in the wrong vacuum. In the hashed region, above and to the right of the dashed black line, the contraint in eq.~\eqref{BackreactionConstraint} is not fulfilled and the backreaction of the Goldberger-Wise scalar on the geometry can not be neglected. The blue, orange, green, red dashed-dotted lines correspond to the radion mass being (neglecting the backreaction, see the discussion at the end of this section) $m_{\rm radion} =  200\, \text{GeV}, 600\, \text{GeV}, 1\, \text{TeV}, 1.4\, \text{TeV}$, respectively.\footnote{The radion mass is, even if the QCD effect is included, dominated by the Goldberger-Wise potential and given by
$$ 
m_{\rm radion}^2 \, \simeq \,  \epsilon \, \frac{4 \pi^2}{3 N^2}  \, \left( v_\ir^2 (4+\epsilon)  \left(\sqrt{\epsilon \, (4 + \epsilon) - \frac{4 \delta}{v_\ir^2}} \, + \, \epsilon \right) \right) \, \mu_{\rm min}^2 \, .
$$
\label{RadionMass}}

The bubble action can also be estimated analytically. For $O(4)$-symmetric bubbles, we are interested in the action at zero temperature, in which case the energy difference between the false and true minima is big and the thick-wall approximation is applicable. This gives \cite{Linde:1981zj}
\be 
\label{S4approx}
S_4 \, \approx \,  \frac{9\, N^4}{8\, \pi^2} \frac{\mu_r^4}{-V_{\rm GW}(\mu_r)} \,.
\ee
The point $\mu_r$ to which the field tunnels is determined by minimizing $S_4$ with respect to $\mu_r$. In fig.~\ref{fig:Resultsna0}, the region where the resulting $S_4$ satisfies eq.~\eqref{PhaseTransitionBubbleAction} and the phase transition thus completes is above the dashed purple line. As one can see, the results using the analytical estimate agree reasonably well with the (more precise) numerical calculation.

Let us next consider $O(3)$-symmetric bubbles. The bounce equation and bubble action are obtained from eqs.~\eqref{BounceEquation} and \eqref{S4Integral} via the replacements $3/r\rightarrow 2/r$, $S_4 \rightarrow S_3$, and ${2 \pi^2 r^3 \rightarrow 4 \pi r^2}$. 
The contribution to the radion potential from the QCD condensates that we include in the next section, requires the temperature to drop below the QCD scale. The action of $O(3)$-symmetric bubbles is then very large and they become unimportant. Since this scenario is the main topic of this paper, we will not calculate their action numerically. At higher temperatures, bubbles with $O(3)$ symmetry can be important though. In order to check if they open up parameter space for the phase transition to complete, we will use an analytical estimate for their action. 
As shown in \cite{Randall:2006py}, the thick-wall approximation is again applicable for these bubbles in the Randall-Sundrum model. The action can then be estimated as \cite{Anderson:1991zb}\footnote{Note that the AdS-Schwarzschild part of the instanton has been neglected in this estimate. Its contribution to the bubble action can not be properly calculated since the normalization of the kinetic term for the field $T_H$ is not known. However, it was argued in ref.~\cite{Creminelli:2001th} that this part of the instanton is suppressed by powers of $N$ relative to the contribution from the Randall-Sundrum space and can therefore justifiably be neglected.}
\be 
S_3 \, \approx \, \frac{\sqrt{3}}{\pi^2} \frac{N^3 \mu_r^3}{\sqrt{V_{\rm GW}(\mu_{\rm min})(T/T_c)^4-V_{\rm GW}(\mu_r)}} \, .
\ee
The term in the denominator arises from the energy difference between the false vacuum (eq.~\eqref{AdS-S-Potential} at $T_H=T$) and the potential at the release point (eq.~\eqref{RadionPotential2} at $\mu=\mu_r$). In fig.~\ref{fig:Resultsna0}, the region where the resulting action satisfies eq.~\eqref{PhaseTransitionBubbleAction} and the phase transition thus completes is above the dotted purple line.
We see that $O(3)$-symmetric bubbles do not open up much more parameter space than those with $O(4)$ symmetry.

Notice that the regions where we found sufficiently small bubble actions are entirely within the hashed regions. Since our calculation of the bubble actions effectively assumes negligible backreaction, the results in the hashed regions are a priori not reliable. In addition, significant backreaction is expected to raise the radion mass to the IR scale, an effect not included in the contour lines. If the radion is so heavy, however, the description of the phase transition in terms of only the radion becomes questionable. Instead one would have to include higher KK modes or calculate the full 5D instanton \cite{Creminelli:2001th,Nardini:2007me}.
It is therefore not clear whether the phase transition can really complete in the corresponding regions of fig.~\ref{fig:Resultsna0}. Furthermore, note that $N=4.5$ is just above the constraint in eq.~\eqref{Nconstraint} from neglecting higher powers of the Ricci scalar in the action. If one increases $N$, these region quickly disappears since the action scales\footnote{This follows from the transformation properties under scale transformations $x \rightarrow \lambda x$ of the kinetic term and potential in eq.~\eqref{S4Integral}. Denoting them by $T$ and $V$, respectively, one has $T \rightarrow \lambda^2 T$ and $V\rightarrow \lambda^4 V$. For $\lambda = N$, we then see that $S_4 \propto N^4$. From the fact that the bounce is an extremum of the action, one can similarly show that $T=-2V$ (see e.g.~ref.\cite{Dasgupta:1996qu}). We use this property as a quality control for the numerical calculation of the bounce.} as $N^4$.

We thus conclude that when using the Goldberger-Wise potential, the RS phase transition does not complete for most of the parameter space. 
Several approaches have been proposed to remedy this situation, for example by deforming the RS geometry \cite{Hassanain:2007js,Konstandin:2010cd,Bunk:2017fic} or by invoking brane-localised curvature \cite{Dillon:2017ctw}.
We show in the next section how QCD confinement provides a universal and effective solution to this problem.

\section{Effect of the QCD condensates on the phase transition}
\label{sec:QCDeffect}

We now include the effect of the QCD condensates on the radion potential and the phase transition. As we have seen in sec.~\ref{sec:QCDcont}, the contribution from the gluon condensate in particular can partly remove the barrier in the Goldberger-Wise potential. We expect that this increases the tunneling rate. Let us now see whether this is the case. 

In regions of parameter space where the phase transition does not complete for the Goldberger-Wise potential, the radion remains stuck in the wrong vacuum. This vacuum has a large, positive cosmological constant since the energy density in the true minimum needs to equal the cosmological constant today and thus almost vanishes (which is achieved by adding a constant contribution to the radion potential). The universe therefore quickly enters an inflationary phase. During this period, the temperature drops exponentially and eventually reaches the QCD scale. The QCD condensates then form and contribute to the radion potential. However, the QCD scale is itself a function of the radion. It decreases with $\mu$ and then saturates at $\Lambda_{\rm QCD}(\mu_c)$ given in eqs.~\eqref{muccondition} and \eqref{muc}. The effect of the QCD condensates on the radion potential is maximized if the temperature drops below this scale $\Lambda_{\rm QCD}(\mu_c)$.
For points in parameter space where the QCD effect allows the phase transition to complete, this provides a lower bound on the temperature $T_n$ at which the phase transition happens:
\be 
T_n \, \gtrsim \Lambda_{\rm QCD}(\mu_c) \, . 
\ee
We emphasize that this is only a lower bound though. For $n=0.1 -0.5$, this then gives
\be
\label{Tnminrange}
T_n^{\rm min} \, \sim \, 10^{-2} \, \text{GeV} \, - \,  10^{-6}  \, \text{GeV}.
\ee
The inflationary phase ends once the phase transition completes. We can estimate the resulting maximal number of e-folds of this stage as 
\be
N_e^{\rm max} \, \sim \, \log \left(\frac{T_c}{T_n^{\rm min}}\right) \, \sim \, \log \left(\frac{\mu_{\rm min}}{T_n^{\rm min}} \right) \, \sim \, 10 \, - \, 20 \, .
\ee
We assume that an earlier stage of inflation is responsible for the features of the Cosmic Microwave Background. A second stage of inflation with a number of e-folds in the above range is then safe from observational constraints.

From eq.~\eqref{Tnminrange}, we expect the phase transition to happen at temperatures far below the critical temperature (for example even for $\epsilon=10^{-3}$, $v_\ir=10^{-2}$ and $N=10$, the latter is of order 10 GeV). Then $O(3)$-symmetric bubbles are highly suppressed and only bubbles with $O(4)$ symmetry are relevant. As discussed in sec.~\ref{sec:QCDcont}, we assume that $n<1$ and restrict ourselves to the gluon condensate and the direction $(\mu,\langle H \rangle =0)$ in the combined radion-Higgs potential. The bounce equation, action and approximate action for $O(4)$-symmetric bubbles is obtained from eqs.~\eqref{BounceEquation} to \eqref{S4approx} by replacing the Goldberger-Wise potential with
\be 
V_{\rm radion}(\mu) \, = \, -V_{\rm QCD}(0) \, + \, V_{\rm QCD}(\mu) \, + \, V_{\rm GW}(\mu) \, .
\ee
The first term is chosen such that the potential vanishes for $\mu=0$. This is an underlying assumption for the bounce equation (which for the Goldberger-Wise potential is automatically fulfilled).

Let us begin with some analytical estimates. The gluon condensate can affect the barrier if its contribution to the potential at the position of the barrier is larger than the barrier height, 
\be
\label{CondensateEffectRegion}
|V_{\rm QCD}(\mu_{\rm max})|\,  \gtrsim \, |V_{\rm GW}(\mu_{\rm max})| \, .
\ee
The importance of the QCD potential relative to the Goldberger-Wise potential grows with decreasing $\mu$. There is then a region $0 \leq \mu \leq \mu_{\rm QCD}$ with some $\mu_{\rm QCD} \gtrsim \mu_{\rm max}$, where we can neglect the Goldberger-Wise potential.  
Using eqs.~\eqref{muc} and \eqref{NewContributionRadionPotential2} (with $\Lambda_{\rm QCD,SM}$ replaced by $\Lambda_{\rm QCD,SM}$ as discussed in sec.~\ref{sec:QCDcont}), the radion potential in this region can be rewritten as
\be 
\label{ApproximatePotential}
V_{\rm radion}(\mu) \, \approx \, \frac{7}{17} \, \Theta(\mu-\mu_c) \, \left(\Lambda_{\rm QCD}(\mu_c) \right)^4 \left( 1 \, - \, \left(\frac{n_c \, \mu}{\Lambda_{\rm QCD}(\mu_c)}\right)^{4n} \right) \, .
\ee
Defining $\tilde{\mu} \equiv \mu / \Lambda_{\rm QCD}(\mu_c)$, the approximate analytical result for the bubble action in eq.~\eqref{S4approx} then gives
\be 
S_4 \, \approx \,  N^4 \, \frac{3}{\pi^2} \, \frac{\tilde{\mu}_r^4}{\Theta(\tilde{\mu}_r - 1/n_c) \, \left( (n_c \, \tilde{\mu}_r)^{4n}  - 1\right)} \, .
\ee
In order to find the bubble action, this expression needs to be minimised with respect to the release point $\tilde{\mu}_r$. As we have discussed in sec.~\ref{sec:QCDIR}, we expect that $n_c \sim \pi$. For values of $n_c$ in the vicinity of this and $n=0.1$, we then find
\be 
\label{S4approxn01}
S_4 \, \approx \, N^4 
\begin{cases} 
0.5 &\quad \text{for} \, \, n_c=2 \\
0.1 &\quad \text{for} \, \, n_c=3 \\
0.03  &\quad \text{for} \, \, n_c=4 \, .    
\end{cases}             
\ee
For $n=0.5$, on the other hand, we find
\be 
\label{S4approxn05}
S_4 \, \approx \, N^4 
\begin{cases} 
0.08 &\quad \text{for} \, \, n_c=2 \\
0.02 &\quad \text{for} \, \, n_c=3 \\
5 \cdot 10^{-3}  &\quad \text{for} \, \, n_c=4 \, .    
\end{cases}             
\ee
The criterion in eq.~\eqref{PhaseTransitionBubbleAction} for the phase transition to complete evaluates to $S_4 \lesssim 50$ for $n=0.1$ and $S_4 \lesssim 90$ for $n=0.5$. We then expect that the phase transition can complete in the entire region delimited by eq.~\eqref{CondensateEffectRegion} for $n=0.1,n_c\gtrsim 3$ or $n=0.5,n_c\gtrsim 2$ and $N=4.5$ (close to its minimal allowed value according to eq.~\eqref{Nconstraint}). 
Depending on the parameters, higher values of $N$ can be viable. For example for $n=0.5$ and $n_c = 4$, the phase transition could complete for $N$ up to $12$. 

\begin{figure}[t]
\centering
\includegraphics[width=7.8cm]{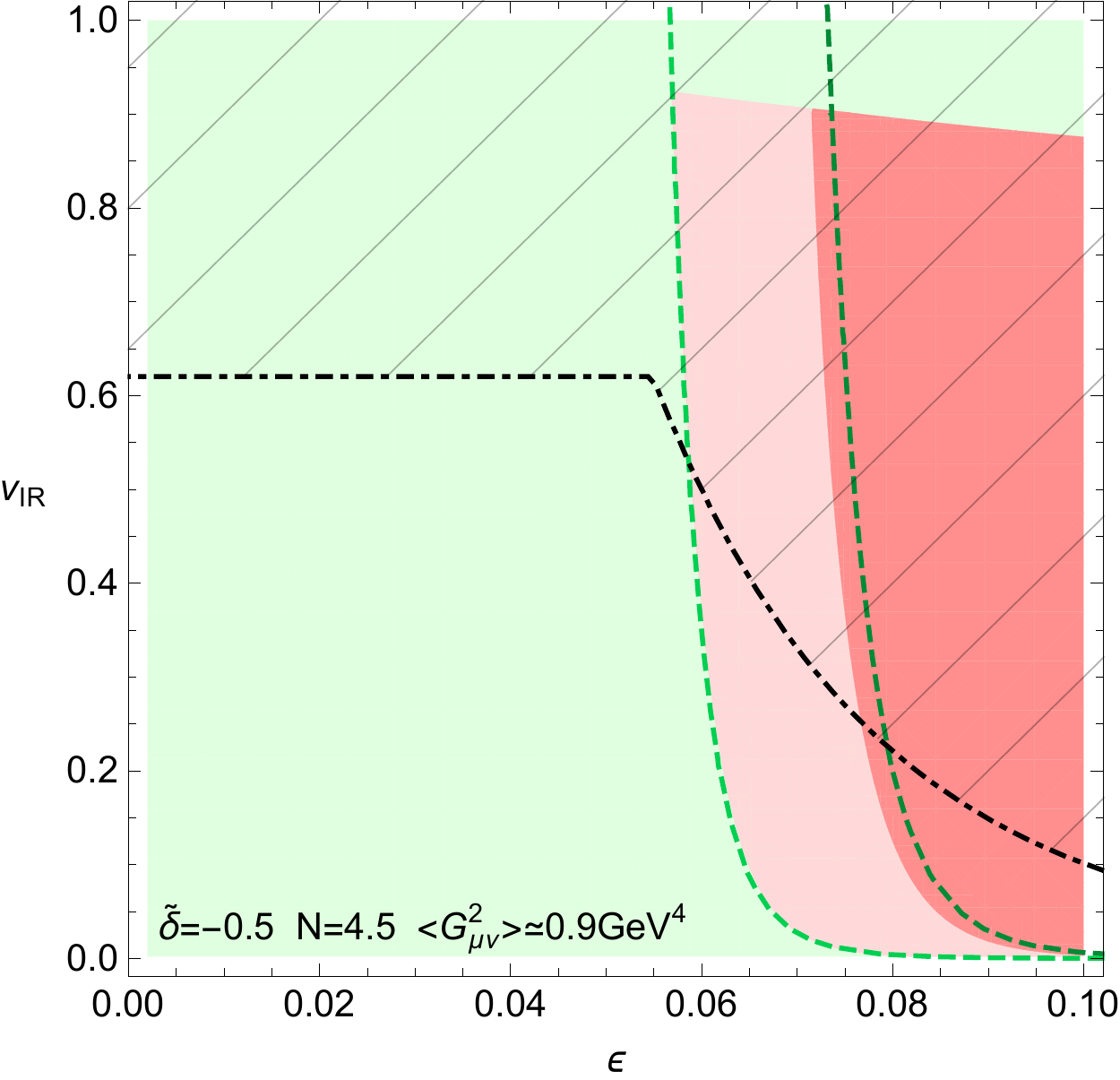}
\hspace{.3cm}
\includegraphics[width=7.8cm]{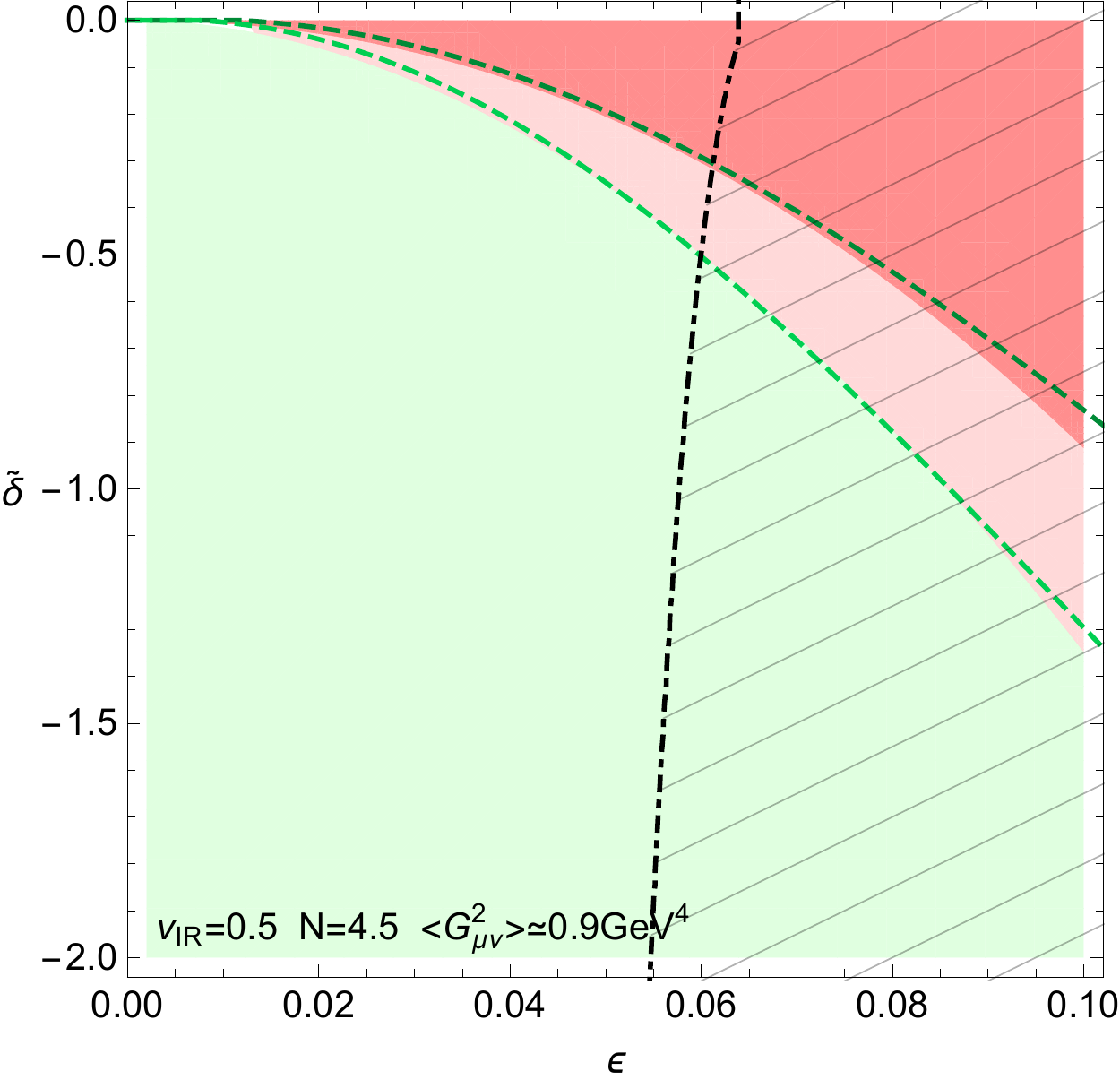}
\caption{\label{fig:Resultsna04} \small 
Results for $\mu_{\rm min }=2.5 \, \text{TeV}$, $N=4.5$ and $n=0.1$ and $n=0.3$ when the QCD effect is included. For the left panel, we have fixed $\tilde{\delta}=-0.5$, and for the right panel, $v_\ir =0.5$. 
Regions where the phase transition can complete via the nucleation of $O(4)$-symmetric bubbles for both $n=0.1$ and $n=0.3$ are shown in green.
This to be compared with the allowed regions in fig.~\ref{fig:Resultsna0} without the QCD effect.
Regions where the nucleation rate is too low are colored in pale (dark) red for $n=0.3$ ($n=0.1$). The corresponding green dashed lines delimit the region satisfying eq.~\eqref{CondensateEffectRegion}, where we expect the QCD effect to be important.
In the hashed region (above the black, dashed line in the left panel and to right of the black, dashed line in the right panel), the backreaction constraint is not fulfilled. The radion masses are as in fig.~\ref{fig:Resultsna0}.
}
\end{figure}

In order to see if these expectations are borne out, we have calculated the action for $O(4)$-symmetric bubbles numerically.
We model $\Lambda_{\rm QCD}(\mu)$ by the function\footnote{Alternatively, one can for example approximate the $\Theta$-function in eq.~\eqref{ApproximatePotential} by $[1+\tanh\left((\mu-\mu_c)/b\right)]/2$ with $b\ll 1$. This is numerically less stable than the choice in eq.~\eqref{LambdaApprox} but gives comparable results.} 
\be 
\label{LambdaApprox}
\Lambda_{\rm QCD}^{\rm approx.}(\mu) \, = \, \Lambda_{\rm QCD,0} \left( \frac{\mu \, e^{-\left(\frac{\mu_c}{\mu}\right)^2} \, + \, \mu_c}{\mu_{\rm min}} \right)^n 
\ee
which smoothly interpolates between eqs.~\eqref{LambdatildeQCD} and \eqref{LambdatildeQCD2} and assume that $n_c=3$. For parameters for which the condition in eq.~\eqref{CondensateEffectRegion} is not satisfied, the gluon condensate typically leads to a second local minimum in the radion potential in the region $\mu < \mu_{\rm max}$. Since the bubble action for tunneling into this local minimum is much smaller than for tunneling directly into the global minimum, the phase transition will then happen in a two-step process.
For parameter points for which such a second minimum appears, we therefore calculate the bubble action for the tunneling from the false vacuum into the local minimum and from there into the global minimum separately. The phase transition can then complete if each of these two bubble actions satisfies the criterion in eq.~\eqref{PhaseTransitionBubbleAction}.

We have performed the calculation for $\mu_{\rm min}=2.5\,$TeV, $N=4.5$, $n=0.3$ and $n=0.1$ and different values of $\epsilon$, $v_\ir$ and $\delta$. As in sec.~\ref{sec:PT}, we choose the parametrization $\delta = \tilde{\delta} \, v_\ir^2$ to ensure that $\delta$ stays in the window above eq.~\eqref{GWminmax} for varying $v_\ir$. In the left panel of fig.~\ref{fig:Resultsna04}, we then fix $\tilde{\delta} = -0.5$ and show results in the $(v_\ir-\epsilon)$-plane. In the right panel, $v_\ir = 0.5$ and results are plotted in the $(\tilde{\delta}-\epsilon)$-plane.
We color the region where the phase transition does not complete for both $n=0.1$ and $n=0.3$ in dark red. The region which is in addition excluded for $n=0.3$ is shaded in pale red, while the remaining allowed region is in green. In the hashed region (above and to the right of the black, dashed line), the contraint in eq.~\eqref{BackreactionConstraint} is again not fulfilled and the backreaction of the Goldberger-Wise scalar on the geometry can not be neglected. Comparing with fig.~\ref{fig:Resultsna0} without the QCD effect, we see that the latter opens up a large region of parameter space. 
We show the regions delimited by the condition in eq.~\eqref{CondensateEffectRegion}, where we expect the gluon condensate to have an effect, as green dashed lines. 
As one can see, these match very well the region where the gluon condensate allows the phase transition to complete. 
To avoid clutter, we have not plotted contour lines for the radion mass. But since the QCD condensates contribute negligibly to the potential near the minimum, they are as in fig.~\ref{fig:Resultsna0}.

\section{Cosmological implications and experimental tests}
\label{sec:cosmoimpli}

In our setup, the cosmological history is the following:  We start at high temperatures in a hot CFT gas. The universe is trapped in the false vacuum at $\mu=0$ separated by a barrier from the true vacuum until the dilaton tunnels out and gets a VEV. This leads to the confinement of the CFT, which induces EW symmetry breaking due to the Higgs-dilaton 
coupling.\footnote{How this Higgs-dilaton interplay happens  depends on the details of the UV completion. From localising the Higgs on the IR brane with a mexican hat potential, a Higgs-dilaton potential of the form $\lambda(H^2-\xi \mu^2)^2/4$ is obtained, where $\xi=v_{\rm EW}^2/\mu^2_{ \rm min}$ and $v_{\rm EW}$ is the Higgs VEV today (see, however, footnote \ref{HiggsMassTermFootnote}). If the Higgs is a pseudo-Nambu-Goldstone boson, the potential instead has the form  $\mu^4(\alpha_0 \sin^2[h/\mu] + \beta_0 \sin^4[h/\mu])$, whose coefficients  depend on the matter content in the composite sector \cite{higgsdilaton}.}  The shallow dilaton potential is associated with a large bubble action, thus a small tunnelling probability, and the universe supercools to very low temperatures. When ignoring QCD effects, one finds that the universe typically remains stuck in the wrong vacuum. We have found that the QCD condensates, on the other hand, can have a large impact on the tunneling action in the radion potential, 
and can enable the RS phase transition to complete in large regions of parameter space.

After tunneling to the release point, the radion starts classically rolling down its potential towards the minimum. 
When the QCD condensates are important, this release point is given by $\mu_r \sim \Lambda_{\rm QCD}(\mu_c)$ which is much smaller than $\mu_{\rm min}$. Since the potential is rather flat for small field values, the field moves slowly and we have to check that quantum fluctuations do not bring it back towards the origin, which would lead to eternal inflation.
To this end, let us consider the equation of motion for $\mu$, 
\be
\ddot{\mu} \, + \, 3 H \dot{\mu} \, = \, \frac{1}{C^3}\frac{\partial V_{\rm radion}}{\partial \mu} \, ,
\ee
where $H$ is the Hubble rate and $C=3N^2/(2\pi^2)$ accounts for the normalisation of the kinetic term of $\mu$.
The quantum fluctuations of the radion in the Hubble background are $\Delta \mu_{\rm quant} \sim H / 2 \pi$ and its classical displacement during one Hubble time $\Delta t \sim H^{-1}$ is 
\be
\Delta \mu_{\rm class} \, \sim \, \Delta t \times \dot{\mu} \, \sim \, \frac{V^{\prime}_{\rm radion}}{3 C^3 H^2} \, , 
\ee
where  we have neglected the $\ddot{\mu}$-term in the equation of motion. We can then define a critical field value $\mu_*$ \cite{Guth:2007ng} for which 
\be
\Delta \mu_{\rm class} \, = \, \Delta \mu_{\rm quant}   \ \ \ \to \ \ \ 
\left. \frac{\partial V_{\rm radion}}{\partial \mu} \right |_{\mu_*}  = \, \frac{3 C^3 H^3}{2 \pi} \, .
\ee
Since $\Delta \mu_{\rm class}$ decreases with $\mu$, quantum fluctuations would dominate the evolution of $\mu$ if $\mu_r \lesssim \mu_*$. We would then be in the regime of eternal inflation. The Hubble rate $H$ is controlled by the vacuum energy in the false vacuum which is approximately of order $\mu^4_{\rm min}$.
This gives
\be
\mu_{*} \, \sim \, \frac{\mu^2_{\rm min}}{M_{\rm Pl}} \, \sim \, \frac{\mbox{TeV}^2}{M_{\rm Pl}} \, \sim \, 10^{-12} \mbox{  GeV} \, .
\ee
For the parameter space of interest, $\mu_r \sim \Lambda_{\rm QCD}(\mu_c)$ is always significantly bigger than this $\mu_*$. 
There is therefore no danger of eternal inflation and the field instead classically rolls towards the minimum of its potential after tunneling. A few e-folds of inflation may result from this short slow-rolling stage. The associated cosmological implications will be discussed in future work \cite{Krpoun}.

The dilaton phase transition can thus trigger the EW  phase transition, and the dynamics that we have studied are therefore directly relevant for EW baryogenesis. Furthermore, since the QCD phase transition takes place before the EW phase transition when 
the top quark is still massless, the QCD phase diagram may be impacted. This potentially makes the QCD phase transition first-order \cite{Pisarski:1983ms,Iso:2017uuu}. Just before the EW phase transition, the temperature of the universe is below the QCD scale. However, the energy density of the universe is dominated by the TeV scale vacuum energy of the dilaton and the universe is inflating. 
After the dilaton phase transition proceeds, reheating takes place as the dilaton energy density is transferred to the standard model particles. We can thus expect that the universe reaches EW scale temperatures again so that the QCD phase transition eventually happens a second time, in the standard way, and the usual standard  thermal history follows. The only remnant of the supercooling stage will be in the form of a stochastic background of gravitational waves observable at LISA \cite{Randall:2006py,Caprini:2015zlo}.
Because of a potentially first-order QCD phase transition preceeding the EW phase transition and both happening when the energy density of the universe is at the TeV scale, this could lead to an enhanced amplitude of the  signal at LISA and potentially some features in the gravitational-wave spectrum (the double-bump would be difficult to resolve though).

We assume a cosmological scenario in which inflation took place at high scales, when the power of density perturbations at the origin of the Cosmic Microwave Background (CMB) was produced, followed  by reheating to high (above the TeV scale) temperatures. The subsequent later stage of supercooling leading to additional 10 to 20 e-folds of inflation may dilute pre-existing heavy particles and potentially have an impact on the axion abundance \cite{Servant:2014bla}. 
A second stage of inflation with a number of e-folds in this range is not constrained, though, and the CMB remains unaffected.

Cosmological implications of this scenario were discussed in \cite{Konstandin:2011dr}, such as the possibility of dilution of relic abundances of stable particles during the TeV-scale inflationary stage or non-thermal dark-matter production during bubble collisions.
A particularly interesting consequence is  baryogenesis. Since the EW phase transition takes place essentially in vacuum, the usual  charge transport mechanism in the vicinity of bubble walls is not appropriate.
This setup is, on the other hand, a natural framework for implementing the mechanism of cold baryogenesis \cite{Servant:2014bla,Konstandin:2011ds} which is very different from the usual EW baryogenesis mechanism and does not involve sphalerons. Instead, Higgs quenching induces Chern-Simons transitions \cite{Tranberg:2006dg,Mou:2017xbo}. In particular, we provide a natural explanation for a nucleation temperature at the QCD scale, which nicely motivates  the possibility that the QCD axion could be responsible for providing enough $CP$-violation for baryogenesis \cite{Servant:2014bla}.

Our scenario works if the dilaton is rather light (below the mass scale of the composite resonances, 
see fig.~\ref{fig:Resultsna0}).
Experimental tests will therefore come from the detection of the dilaton.
In a natural realisation without fine-tuning, this means the dilaton is accessible at the LHC. 
The properties and signatures of the dilaton are determined by its Nambu-Goldstone nature \cite{Goldberger:2008zz,Bellazzini:2012vz,Chacko:2012vm,Blum:2014jca,Megias:2015qqh} and it can be distinguished from other additional singlets in extended scalar sectors of the standard model. While discovering a light dilaton at the LHC would be a signal in favour of our scenario, this would not be enough. In fact, as clearly shown by the comparison between figs.~\ref{fig:Resultsna0} and \ref{fig:Resultsna04}, we also need a small parameter $n$ for the phase transition to complete. According to eqs.~(\ref{eq:bcft}) and (\ref{eq:nversusbeta}), such a small $n$ corresponds to a small $|b_{CFT}|$ which means a large 5D gauge coupling $k g_5^2$. This coupling can be probed by measuring resonant production of KK gluons or KK quarks.
Given the  bounds from EW precision tests of order 2 TeV on the KK scale (see footnote 1), we conclude that a future high energy collider is needed to probe this scenario.

\section{Conclusions}
\label{sec:conc}
Our analysis shows that the possibility to delay the electroweak phase transition down to QCD temperatures can arise naturally in models where electroweak symmetry breaking is linked to nearly conformal sectors, as is well-motivated in Randall-Sundrum and composite Higgs models. 
We have found that the first-order Randall-Sundrum phase transition becomes much more likely  when including effects from the QCD condensate in the radion/dilaton potential. 
The comparison between figs.~\ref{fig:Resultsna0} and \ref{fig:Resultsna04} shows that a large region of parameter space opens up when incorporating this effect in the analysis of the tunneling probability.
We summarise our key ingredients:
\begin{itemize}
\item The dilaton VEV $\mu$ determines the confinement scale of the CFT (or the radion VEV the IR scale of the RS model).
\item The Higgs acquires a potential from its coupling to the radion/dilaton and gets a non-zero VEV controlled by the scale $\mu$.

\item The composite sector is colored and thus coupled to gluons, as imposed by the scenario of 
 composite Higgs models with partial compositeness \cite{Panico:2015jxa,Contino:2006nn}.
At energies lower than $\mu$, the CFT degrees of freedom no longer contribute to the running of the QCD coupling.
The latter thereby depends on $\mu$. In the Randall-Sundrum description, this results from the gluon living in the bulk of the extra dimension.

\item Consequently, the scale $\Lambda_{\rm QCD}$ at which the QCD coupling becomes strong depends on $\mu$ too. It scales as (see eq.~\eqref{LambdatildeQCD})
$$
\Lambda_{\rm QCD}(\mu) \, \propto \,  \mu^n \ \ \ \  \text{for} \, \, \,  \Lambda_{\rm QCD}(\mu)\, \lesssim \, \mu \, .
$$
This means that at small $\mu$, the blowing-up of the QCD coupling is delayed for $n<1$,  where $n$ is a free parameter determined by the CFT degrees of freedom contributing to the
QCD $\beta$-function, or by the 5D gauge coupling (see eq.~(\ref{eq:nversusbeta})).

\item The  QCD contribution from gluon condensation to the radion/dilaton potential comes with a negative sign, (see 
eq.~(\ref{NewContributionRadionPotential})), 
$$
V_{\rm QCD} \, \propto \,  -  \, \left(\Lambda_{\rm QCD}(\mu) \right)^4 \, .
$$
This lowers the potential at small $\mu$ and contributes to remove a significant part of the barrier, leading to an important impact on the tunneling action for $0.1\lesssim n \lesssim 1$.

\end{itemize}

While we have not included the Higgs in our analysis, the Higgs-radion interplay can lead to additional non-trivial 
effects such as Yukawa coupling variation during the electroweak phase transition \cite{vonHarling:2016vhf,higgsdilaton}.

The effect that we have studied in detail in the context of the Randall-Sundrum model is rather general and applies to other nearly conformal potentials.
A related discussion was presented in ref.~\cite{Iso:2017uuu} in the framework of a classically conformal $B - L$-extension of the standard model. In this context, the Higgs has a  quartic coupling to an additional singlet scalar  whose VEV breaks $B-L$. 
The corresponding Coleman-Weinberg potential has no barrier at zero temperature
(it remains to be checked whether in this model a zero-temperature barrier could be generated through slowly running perturbing operators at higher order, similar to what happens in the Goldberger-Wise mechanism).
Rather than the gluon condensate,  the top-quark condensate is considered, which generates a linear term in the Higgs potential  $y_t \langle \bar{t} t\rangle h/\sqrt{2}$ that suppresses the thermal barrier from the Higgs thermal mass at temperatures below the QCD scale.

Our results strongly motivate the possibility of cold baryogenesis \cite{Konstandin:2011ds,Servant:2014bla,Mou:2017xbo} and open novel opportunities for  cosmology such as a natural TeV-scale stage of inflation which can lead to the production of 
primordial black holes \cite{Krpoun}. The  strongly first-order electroweak phase transition also leads to a large stochastic signal of gravitational waves observable at LISA  \cite{Randall:2006py,Caprini:2015zlo}.

\section*{Acknowledgments}
We thank Kaustubh Agashe, Iason Baldes, Valerie Domcke, Stephan Huber, Thomas Konstandin, Christoph Krpoun, Kyohei Mukaida, Oriol Pujolas, Pasquale Serpico, Kengo Shimada and  Alexander Westphal for discussions. We acknowledge the stimulating participation of the DESY summer students Marc Basquens Munoz and Hans Joos
during the early stages of this work in summer 2016. BvH thanks Fermilab for hospitality while part of this work was completed. This visit has received funding/support from the European Union's Horizon 2020 research and innovation programme under the Marie Sk\l{}odowska-Curie grant agreement No 690575.

\section*{Appendix: Perturbativity constraints for Goldberger-Wise}

In order to be able to neglect the backreaction of the Goldberger-Wise scalar on the geometry, the magnitude of its energy-momentum tensor needs to be smaller than the contribution from the cosmological constant. The energy momentum tensor of the Goldberger-Wise scalar is given by
\be 
T^{MN}_\phi \, = \, g^{MN} \left(\frac{1}{2} \left(\partial_M  \phi \right)^2 \, - \, \frac{m^2}{2} \phi^2\right)\, - \, \left(\partial^M \phi \right) \left(\partial^N \phi \right) \, .
\ee
Plugging in its VEV in eq.~\eqref{vevprofile}, the $\mu\nu$-components read
\begin{multline} 
\label{EMTGW}
T^{\mu \nu}_\phi \, \simeq \, k^5 v_\ir^2 \eta^{\mu \nu } \Bigl(\frac{k}{z}\Bigr)^2 \left[(8 +2 \epsilon)\left(1-\Bigl(\frac{\mu}{\mu_{\rm min}}\Bigr)^\epsilon X_{\rm min} \right)^2 \Bigl(\frac{\mu}{z} \Bigr)^{8+2\epsilon} -  \, 2 \, \epsilon \, \Bigl(\frac{z}{\mu_{\rm min}}\Bigr)^{2\epsilon} X_{\rm min}^2 \right. \\ \left.  - \, 8 \, \epsilon \, \left(1-\Bigl(\frac{\mu}{\mu_{\rm min}}\Bigr)^\epsilon X_{\rm min} \right) \Bigl(\frac{\mu}{z} \Bigr)^{4+\epsilon} \Bigl(\frac{z}{\mu_{\rm min}} \Bigr)^\epsilon X_{\rm min}\right] \, ,
\end{multline}
where $z\equiv k e^{-k y}$. The $55$-component leads to the same constraints and will therefore not be given explicitly. The energy-momentum tensor due to the bulk cosmological constant in the RS model is given by
\be 
\label{EMTLambda}
T^{MN}_\Lambda \, = \, - 24 \, M_5^3 \, k^2 g^{MN} .
\ee
We need to impose that the absolute value of eq.~\eqref{EMTGW} is smaller than that of eq.~\eqref{EMTLambda} everywhere in the bulk, $k \geq z \geq \mu$, and for different radion values $\mu$. For the analysis of the phase transition, we are in particular interested in the radion potential for the range $0 \leq \mu \lesssim \mu_{\rm min}$. For $\epsilon>0$, the VEV of the Goldberger-Wise scalar decreases in magnitude when going from both the IR and UV brane towards the bulk. It is then sufficient to compare the energy-momentum tensors near the two branes. From the UV brane, this gives the constraint
\be
v_\ir \, \ll \, \frac{N }{2 \pi X_{\rm min}} \sqrt{\frac{3}{\epsilon}} \, \Bigl(\frac{\mu_{\rm min}}{k}\Bigr)^\epsilon \, .
\ee
Near the IR brane, on the other hand, the energy-momentum tensor of the Goldberger-Wise scalar depends more strongly on the radion VEV $\mu$. The most stringent constraint in the range of interest $0 \leq \mu \lesssim \mu_{\rm min}$ arises for $\mu \ll \mu_{\rm min}$, where terms in eq.~\eqref{EMTGW} involving $(\mu/\mu_{\rm min})^\epsilon$ can be neglected. This gives
\be 
v_\ir \, \ll \, \frac{\sqrt{3} N }{4 \pi } \, .
\ee
Note that in the literature, the constraint has instead sometimes been evaluated for $\mu=\mu_{\rm min}$. This leads to the factors $(1-(\mu/\mu_{\rm min})^\epsilon X_{\rm min})$ in eq.~\eqref{EMTGW} being suppressed by $\epsilon$. The resulting less stringent constraint can then not guarantee that the backreaction can be neglected for $\mu \ll \mu_{\rm min}$.

For $\epsilon <0$, the terms involving $(\mu/\mu_{\rm min})^\epsilon$ become arbitrarily big for sufficiently small $\mu$. In this case, there is therefore always a region around the origin in the radion potential for which the backreaction can not be neglected.\footnote{This has been noticed already e.g.~in ref.~\cite{DeWolfe:1999cp}.}


\begin{thebibliography}{99}


\bibitem{Caprini:2015zlo} 
  C.~Caprini {\it et al.},
  ``Science with the space-based interferometer eLISA. II: Gravitational waves from cosmological phase transitions,''
  JCAP {\bf 1604}, no. 04, 001 (2016)
  [arXiv:1512.06239 [astro-ph.CO]].


\bibitem{Witten:1980ez} 
  E.~Witten,
  ``Cosmological Consequences of a Light Higgs Boson,''
  Nucl.\ Phys.\ B {\bf 177}, 477 (1981).
  
\bibitem{Buchmuller:1990ds}
  W.~Buchmuller and D.~Wyler,
  ``The Effect of dilatons on the electroweak phase transition,''
  Phys.\ Lett.\ B {\bf 249} (1990) 281.

  
\bibitem{Espinosa:2008kw}
  J.~R.~Espinosa, T.~Konstandin, J.~M.~No and M.~Quiros,
  ``Some Cosmological Implications of Hidden Sectors,''
  Phys.\ Rev.\ D {\bf 78} (2008) 123528
  [arXiv:0809.3215 [hep-ph]].
  
\bibitem{Konstandin:2011dr} 
  T.~Konstandin and G.~Servant,
  ``Cosmological Consequences of Nearly Conformal Dynamics at the TeV scale,''
  JCAP {\bf 1112}, 009 (2011)
  [arXiv:1104.4791 [hep-ph]].

  
\bibitem{Jaeckel:2016jlh}
  J.~Jaeckel, V.~V.~Khoze and M.~Spannowsky,
  ``Hearing the signals of dark sectors with gravitational wave detectors,''
  Phys.\ Rev.\ D {\bf 94} (2016) no.10,  103519
  [arXiv:1602.03901 [hep-ph]].
   
 
\bibitem{Marzola:2017jzl}
  L.~Marzola, A.~Racioppi and V.~Vaskonen,
  ``Phase transition and gravitational wave phenomenology of scalar conformal extensions of the Standard Model,''
  Eur.\ Phys.\ J.\ C {\bf 77} (2017) no.7,  484
  [arXiv:1704.01034 [hep-ph]].
  
\bibitem{Iso:2017uuu} 
  S.~Iso, P.~D.~Serpico and K.~Shimada,
  ``QCD-Electroweak First-Order Phase Transition in a Supercooled Universe,''
  Phys.\ Rev.\ Lett.\  {\bf 119}, no. 14, 141301 (2017)
  [arXiv:1704.04955 [hep-ph]].

\bibitem{Servant:2014bla} 
  G.~Servant,
  ``Baryogenesis from Strong $CP$ Violation and the QCD Axion,''
  Phys.\ Rev.\ Lett.\  {\bf 113}, no. 17, 171803 (2014)
  [arXiv:1407.0030 [hep-ph]].


\bibitem{Randall:1999ee} 
  L.~Randall and R.~Sundrum,
  ``A Large mass hierarchy from a small extra dimension,''
  Phys.\ Rev.\ Lett.\  {\bf 83}, 3370 (1999)
  [hep-ph/9905221].

\bibitem{Agashe:2004rs}
  K.~Agashe, R.~Contino and A.~Pomarol,
  ``The Minimal composite Higgs model,''
  Nucl.\ Phys.\ B {\bf 719} (2005) 165
  [hep-ph/0412089].

\bibitem{Contino:2003ve}
  R.~Contino, Y.~Nomura and A.~Pomarol,
  ``Higgs as a holographic pseudoGoldstone boson,''
  Nucl.\ Phys.\ B {\bf 671} (2003) 148
  [hep-ph/0306259].


\bibitem{Contino:2010rs} 
  R.~Contino,
  ``The Higgs as a Composite Nambu-Goldstone Boson,''
  arXiv:1005.4269 [hep-ph].
  
  
\bibitem{Panico:2015jxa}
  G.~Panico and A.~Wulzer,
  ``The Composite Nambu-Goldstone Higgs,''
  Lect.\ Notes Phys.\  {\bf 913} (2016) pp.1
  [arXiv:1506.01961 [hep-ph]].

\bibitem{Contino:2006nn}
  R.~Contino, T.~Kramer, M.~Son and R.~Sundrum,
  ``Warped/composite phenomenology simplified,''
  JHEP {\bf 0705} (2007) 074
  [hep-ph/0612180].

\bibitem{Gherghetta:2000qt}
  T.~Gherghetta and A.~Pomarol,
  ``Bulk fields and supersymmetry in a slice of AdS,''
  Nucl.\ Phys.\ B {\bf 586} (2000) 141
  [hep-ph/0003129].

\bibitem{Huber:2000ie}
  S.~J.~Huber and Q.~Shafi,
  ``Fermion masses, mixings and proton decay in a Randall-Sundrum model,''
  Phys.\ Lett.\ B {\bf 498} (2001) 256
  [hep-ph/0010195].

\bibitem{Gherghetta:2010cj} 
  T.~Gherghetta,
  ``A Holographic View of Beyond the Standard Model Physics,''
  arXiv:1008.2570 [hep-ph].

\bibitem{Goldberger:1999uk}
  W.~D.~Goldberger and M.~B.~Wise,
  ``Modulus stabilization with bulk fields,''
  Phys.\ Rev.\ Lett.\  {\bf 83} (1999) 4922
  [hep-ph/9907447].

\bibitem{Creminelli:2001th}
  P.~Creminelli, A.~Nicolis and R.~Rattazzi,
  ``Holography and the electroweak phase transition,''
  JHEP {\bf 0203} (2002) 051
  [hep-th/0107141].
  
  \bibitem{Randall:2006py}
  L.~Randall and G.~Servant,
  ``Gravitational waves from warped spacetime,''
  JHEP {\bf 0705} (2007) 054
  [hep-ph/0607158].


\bibitem{Nardini:2007me} 
  G.~Nardini, M.~Quiros and A.~Wulzer,
  ``A Confining Strong First-Order Electroweak Phase Transition,''
  JHEP {\bf 0709}, 077 (2007)
  [arXiv:0706.3388 [hep-ph]].

\bibitem{Hassanain:2007js} 
  B.~Hassanain, J.~March-Russell and M.~Schvellinger,
  ``Warped Deformed Throats have Faster (Electroweak) Phase Transitions,''
  JHEP {\bf 0710}, 089 (2007)
  [arXiv:0708.2060 [hep-th]].



\bibitem{Bunk:2017fic} 
  D.~Bunk, J.~Hubisz and B.~Jain,
 ``A Perturbative RS I Cosmological Phase Transition,''
  arXiv:1705.00001 [hep-ph].

\bibitem{Konstandin:2010cd}
  T.~Konstandin, G.~Nardini and M.~Quiros,
  ``Gravitational Backreaction Effects on the Holographic Phase Transition,''
  Phys.\ Rev.\ D {\bf 82} (2010) 083513
  [arXiv:1007.1468 [hep-ph]].
  

\bibitem{Dillon:2017ctw} 
  B.~M.~Dillon, B.~K.~El-Menoufi, S.~J.~Huber and J.~P.~Manuel,
  ``A rapid holographic phase transition with brane-localized curvature,''
  arXiv:1708.02953 [hep-th].



\bibitem{Breitenlohner:1982bm}
  P.~Breitenlohner and D.~Z.~Freedman,
  ``Positive Energy in anti-De Sitter Backgrounds and Gauged Extended Supergravity,''
  Phys.\ Lett.\  {\bf 115B} (1982) 197.


\bibitem{Agashe:2007zd}
  K.~Agashe, H.~Davoudiasl, G.~Perez and A.~Soni,
  ``Warped Gravitons at the LHC and Beyond,''
  Phys.\ Rev.\ D {\bf 76} (2007) 036006
  [hep-ph/0701186].

\bibitem{Csaki:1999mp}
  C.~Csaki, M.~Graesser, L.~Randall and J.~Terning,
  ``Cosmology of brane models with radion stabilization,''
  Phys.\ Rev.\ D {\bf 62} (2000) 045015
  [hep-ph/9911406].


\bibitem{Goldberger:1999un}
  W.~D.~Goldberger and M.~B.~Wise,
  ``Phenomenology of a stabilized modulus,''
  Phys.\ Lett.\ B {\bf 475} (2000) 275
  [hep-ph/9911457].



\bibitem{Malm:2013jia}
  R.~Malm, M.~Neubert, K.~Novotny and C.~Schmell,
  ``5D Perspective on Higgs Production at the Boundary of a Warped Extra Dimension,''
  JHEP {\bf 1401} (2014) 173
  [arXiv:1303.5702 [hep-ph]].

\bibitem{Bauer:2016lbe}
  M.~Bauer, C.~H\"orner and M.~Neubert,
  ``Diphoton Resonance from a Warped Extra Dimension,''
  JHEP {\bf 1607} (2016) 094
  [arXiv:1603.05978 [hep-ph]].

\bibitem{Bauer:2011ah}
  M.~Bauer, R.~Malm and M.~Neubert,
  ``A Solution to the Flavor Problem of Warped Extra-Dimension Models,''
  Phys.\ Rev.\ Lett.\  {\bf 108} (2012) 081603
  [arXiv:1110.0471 [hep-ph]].

\bibitem{vonHarling:2016vhf} 
  B.~von Harling and G.~Servant,
  ``Cosmological evolution of Yukawa couplings: the 5D perspective,''
  JHEP {\bf 1705}, 077 (2017)
  [arXiv:1612.02447 [hep-ph]].


\bibitem{Csaki:2007ns}
  C.~Csaki, J.~Hubisz and S.~J.~Lee,
  ``Radion phenomenology in realistic warped space models,''
  Phys.\ Rev.\ D {\bf 76} (2007) 125015
  [arXiv:0705.3844 [hep-ph]].
  


\bibitem{Agashe:2002bx}
  K.~Agashe, A.~Delgado and R.~Sundrum,
  ``Gauge coupling renormalization in RS1,''
  Nucl.\ Phys.\ B {\bf 643} (2002) 172
  [hep-ph/0206099].
  
  
  
\bibitem{Goldberger:2003mi} 
  W.~D.~Goldberger and I.~Z.~Rothstein,
  ``Systematics of coupling flows in AdS backgrounds,''
  Phys.\ Rev.\ D {\bf 68}, 125012 (2003)
  [hep-ph/0303158].
  
\bibitem{Goldberger:2002cz} 
  W.~D.~Goldberger and I.~Z.~Rothstein,
  ``High-energy field theory in truncated AdS backgrounds,''
  Phys.\ Rev.\ Lett.\  {\bf 89}, 131601 (2002)
  [hep-th/0204160].
  
\bibitem{Goldberger:2002hb} 
  W.~D.~Goldberger and I.~Z.~Rothstein,
  ``Effective field theory and unification in AdS backgrounds,''
  Phys.\ Rev.\ D {\bf 68}, 125011 (2003)
  [hep-th/0208060].
  
  
  

\bibitem{Narison:2011xe}
  S.~Narison,
  ``Gluon Condensates and precise $\overline{m}_{c,b}$ from QCD-Moments and their ratios to Order $\alpha_s^3$ and $\langle$G$^4 \rangle$,''
  Phys.\ Lett.\ B {\bf 706} (2012) 412
  [arXiv:1105.2922 [hep-ph]].
    
\bibitem{Narison:2010cg}
  S.~Narison,
  ``Gluon condensates and c, b quark masses from quarkonia ratios of moments,''
  Phys.\ Lett.\ B {\bf 693} (2010) 559
   Erratum: [Phys.\ Lett.\ B {\bf 705} (2011) 544]
  [arXiv:1004.5333 [hep-ph]].
    
\bibitem{Narison:2011rn}
  S.~Narison,
  ``Gluon Condensates and $m_b(m_b)$ from QCD-Exponential Moments at Higher Orders,''
  Phys.\ Lett.\ B {\bf 707} (2012) 259
  [arXiv:1105.5070 [hep-ph]].

\bibitem{Ioffe:2005ym}
  B.~L.~Ioffe,
  ``QCD at low energies,''
  Prog.\ Part.\ Nucl.\ Phys.\  {\bf 56} (2006) 232
  [hep-ph/0502148].
  
  
\bibitem{DiGiacomo:1981lcx}
  A.~Di Giacomo and G.~C.~Rossi,
  ``Extracting the Vacuum Expectation Value of the Quantity alpha / pi G G from Gauge Theories on a Lattice,''
  Phys.\ Lett.\  {\bf 100B} (1981) 481.
  
\bibitem{Campostrini:1989uj}
  M.~Campostrini, A.~Di Giacomo and Y.~Gunduc,
  ``Gluon Condensation in SU(3) Lattice Gauge Theory,''
  Phys.\ Lett.\ B {\bf 225} (1989) 393.
  
\bibitem{Rakow:2005yn}
  P.~E.~L.~Rakow,
  ``Stochastic perturbation theory and the gluon condensate,''
  PoS LAT {\bf 2005} (2006) 284
  [hep-lat/0510046].


\bibitem{Patrignani:2016xqp}
  C.~Patrignani {\it et al.} [Particle Data Group],
  Chin.\ Phys.\ C {\bf 40} (2016) no.10,  100001.
  doi:10.1088/1674-1137/40/10/100001
  
  
\bibitem{Shifman:1978bx}
  M.~A.~Shifman, A.~I.~Vainshtein and V.~I.~Zakharov,
  ``QCD and Resonance Physics. Theoretical Foundations,''
  Nucl.\ Phys.\ B {\bf 147} (1979) 385.



\bibitem{Coleman:1977py}
  S.~R.~Coleman,
  ``The Fate of the False Vacuum. 1. Semiclassical Theory,''
  Phys.\ Rev.\ D {\bf 15} (1977) 2929
   Erratum: [Phys.\ Rev.\ D {\bf 16} (1977) 1248].


\bibitem{Callan:1977pt}
  C.~G.~Callan, Jr. and S.~R.~Coleman,
  ``The Fate of the False Vacuum. 2. First Quantum Corrections,''
  Phys.\ Rev.\ D {\bf 16} (1977) 1762.

\bibitem{Linde:1981zj}
  A.~D.~Linde,
  ``Decay of the False Vacuum at Finite Temperature,''
  Nucl.\ Phys.\ B {\bf 216} (1983) 421
   Erratum: [Nucl.\ Phys.\ B {\bf 223} (1983) 544].


\bibitem{Anderson:1991zb}
  G.~W.~Anderson and L.~J.~Hall,
  ``The Electroweak phase transition and baryogenesis,''
  Phys.\ Rev.\ D {\bf 45} (1992) 2685.
  
  
\bibitem{Dasgupta:1996qu}
  I.~Dasgupta,
  ``Estimating vacuum tunneling rates,''
  Phys.\ Lett.\ B {\bf 394} (1997) 116
  [hep-ph/9610403].

  
\bibitem{higgsdilaton}
S.~Bruggisser, B.~von Harling, O.~Matsedonskyi and G.~Servant, 
``Electroweak phase transition in composite Higgs models'',
in preparation.  
  
  
\bibitem{Guth:2007ng} 
  A.~H.~Guth,
  ``Eternal inflation and its implications,''
  J.\ Phys.\ A {\bf 40}, 6811 (2007)
  [hep-th/0702178].

\bibitem{Krpoun}
  C.~Krpoun, G.~Servant and A.~Westphal, in preparation.
  


\bibitem{Pisarski:1983ms} 
  R.~D.~Pisarski and F.~Wilczek,
  ``Remarks on the Chiral Phase Transition in Chromodynamics,''
  Phys.\ Rev.\ D {\bf 29}, 338 (1984).


\bibitem{Konstandin:2011ds} 
  T.~Konstandin and G.~Servant,
  ``Natural Cold Baryogenesis from Strongly Interacting Electroweak Symmetry Breaking,''
  JCAP {\bf 1107}, 024 (2011)
  [arXiv:1104.4793 [hep-ph]].



\bibitem{Mou:2017xbo}
  Z.~G.~Mou, P.~M.~Saffin and A.~Tranberg,
  ``Simulations of Cold Electroweak Baryogenesis: Quench from portal coupling to new singlet field,''
  arXiv:1711.04524 [hep-ph].
  

   
\bibitem{Tranberg:2006dg}
  A.~Tranberg, J.~Smit and M.~Hindmarsh,
  ``Simulations of cold electroweak baryogenesis: Finite time quenches,''
  JHEP {\bf 0701} (2007) 034
  [hep-ph/0610096].

  
\bibitem{Goldberger:2008zz}
  W.~D.~Goldberger, B.~Grinstein and W.~Skiba,
  ``Distinguishing the Higgs boson from the dilaton at the Large Hadron Collider,''
  Phys.\ Rev.\ Lett.\  {\bf 100} (2008) 111802
  [arXiv:0708.1463 [hep-ph]].
  
\bibitem{Bellazzini:2012vz}
  B.~Bellazzini, C.~Csaki, J.~Hubisz, J.~Serra and J.~Terning,
  ``A Higgslike Dilaton,''
  Eur.\ Phys.\ J.\ C {\bf 73} (2013) no.2,  2333
  [arXiv:1209.3299 [hep-ph]].
  
\bibitem{Chacko:2012vm}
  Z.~Chacko, R.~Franceschini and R.~K.~Mishra,
  ``Resonance at 125 GeV: Higgs or Dilaton/Radion?,''
  JHEP {\bf 1304} (2013) 015
  [arXiv:1209.3259 [hep-ph]].
  
\bibitem{Blum:2014jca}
  K.~Blum, M.~Cliche, C.~Csaki and S.~J.~Lee,
  ``WIMP Dark Matter through the Dilaton Portal,''
  JHEP {\bf 1503} (2015) 099
  [arXiv:1410.1873 [hep-ph]].
  
\bibitem{Megias:2015qqh}
  E.~Megias, O.~Pujolas and M.~Quiros,
  ``On light dilaton extensions of the Standard Model,''
  EPJ Web Conf.\  {\bf 126} (2016) 05010
  [arXiv:1512.06702 [hep-ph]].


\bibitem{DeWolfe:1999cp}
  O.~DeWolfe, D.~Z.~Freedman, S.~S.~Gubser and A.~Karch,
  ``Modeling the fifth-dimension with scalars and gravity,''
  Phys.\ Rev.\ D {\bf 62} (2000) 046008
  [hep-th/9909134].
 
 


  
\end{thebibliography}
\end{document}